\documentclass{article}
\pdfoutput=1

\usepackage{arxiv}
\usepackage[utf8]{inputenc} 
\usepackage[T1]{fontenc} 
\usepackage{hyperref} 
\hypersetup{
	colorlinks = true,
	linkcolor = black,
	anchorcolor = black,
	citecolor = black,
	filecolor = black,
	urlcolor = blue
}
\usepackage{booktabs} 
\usepackage{amsfonts} 
\usepackage{nicefrac} 
\usepackage{microtype} 
\usepackage{amsmath,bm}
\usepackage{graphicx}
\usepackage[flushleft]{threeparttable}
\usepackage{multirow}
\usepackage{natbib}
\usepackage{makecell}
\usepackage{caption}
\usepackage{subcaption}   
\usepackage{nicematrix}
\usepackage{float}

\newcommand{\figurehere}[1]{\begin{center}%
		=========================\\%
		Insert Figure #1 about here\\%
		=========================\\%
\end{center}}
\newcommand{\tablehere}[1]{\begin{center}%
		=========================\\%
		Insert Table #1 about here\\%
		=========================\\%
\end{center}}

\usepackage{array}
\newcommand{\PreserveBackslash}[1]{\let\temp=\\#1\let\\=\temp}
\newcolumntype{C}[1]{>{\PreserveBackslash\centering}p{#1}}
\newcolumntype{R}[1]{>{\PreserveBackslash\raggedleft}p{#1}}
\newcolumntype{L}[1]{>{\PreserveBackslash\raggedright}p{#1}}

\title{Further Exploration of the Effects of Time-varying Covariate in Growth Mixture Models with Nonlinear Trajectories}

\author{
  Jin Liu \thanks{CONTACT Jin Liu Email: Veronica.Liu0206@gmail.com}\\
Department of Biostatistics\\
Virginia Commonwealth University 
}

\begin{document}

\maketitle
\begin{abstract}
Growth mixture modeling (GMM) is an analytical tool for identifying multiple unobserved sub-populations of longitudinal processes. In particular, it describes change patterns within each latent sub-population and examines between-individual differences in within-individual change for each sub-group. One research interest in utilizing GMMs is to explore how covariates affect such heterogeneity in change patterns. \citet{Liu2021MoE} extended mixture-of-experts (MoE) models, which mainly focus on time-invariant covariates, to allow the covariates to account for within-group and between-group differences simultaneously and examine the heterogeneity in nonlinear trajectories. The present study further extends \citet{Liu2021MoE} and examines the effects on trajectory heterogeneity of time-varying covariates (TVCs). Specifically, we propose methods to decompose a TVC into a trait feature (e.g., the baseline value of a TVC) and a set of state features (e.g., interval-specific slopes or changes). The trait features are allowed to account for within-group differences in growth factors of trajectories (i.e., trait effect), and the state features are allowed to impact observed values of a longitudinal process (i.e., state effect). We examine the proposed models using a simulation study and a real-world data analysis. The simulation study demonstrated that the proposed models are capable of separating trajectories into several clusters and generally generating unbiased and accurate estimates with target coverage probabilities. With the proposed models, we showed the heterogeneity in the trait and state features of reading ability across latent classes of students' mathematics performance. Meanwhile, the trait and state effects on mathematics development of reading ability are also heterogeneous across the clusters of students.
\end{abstract}


\setcounter{secnumdepth}{3}
\section*{Introduction}\label{Intro}
Growth mixture modeling (GMM) is an analytical tool to examine heterogeneity in longitudinal processes and describe change patterns of trajectories for each identified sub-group. This tool stemmed from a modeling framework named finite mixture modeling (FMM), which was introduced to the structural equation modeling (SEM) family by \citet{Muthen1999GMM}. As its name suggests, the FMM represents sample heterogeneity by allowing for a finite number of unobserved classes and mixing these latent groups using a linear combination. When utilized to examine growth curves, the FMM is then called GMM. Similar to the FMM, the GMM is a model-based clustering method \citep{Bouveyron2019GMM} that enables researchers to analyze data with domain knowledge and construct models to test hypotheses. Moreover, the GMM is a probability-based clustering approach that allows each trajectory to belong to multiple classes simultaneously with different probabilities (i.e., uncertainty), based on which the algorithm decides the cluster each trajectory enters. The GMM has received considerable attention over the past 20 years. Many theoretical and empirical works, such as \citet{Bauer2003GMM, Muthen2004GMM, Grimm2009FMM, Grimm2010FMM, Nylund2007number}, have examined its advantages and disadvantages. 

In addition to analyzing the heterogeneity of processes, it is also of interest to assess how covariates contribute to the between-group and within-group differences in such within-individual processes when utilizing the GMM. \citet{Liu2021MoE} demonstrated that two types of time-invariant covariates (TICs) could be added into the GMM: (1) TICs that suggest group components and (2) TICs that explain variability within each group. Specifically, \citet{Liu2021MoE} proposed building up a full mixture model to simultaneously allow for these two types of TICs in a GMM. With an extensive simulation study, this work examined the model performance and how the two types of TICs affect clustering and estimating effects. Moreover, this research work pointed out that the covariates that account for within-group heterogeneity could also be time-varying and showed it with real-world data analysis. The GMM with a time-varying covariate (TVC) in \citet{Liu2021MoE} allows for the assessment of the heterogeneity in the TVC and its heterogeneous effects on the longitudinal process across latent classes. However, the model has some limitations. First, the model does not allow the TVC to predict variability in the growth factors in each cluster of the longitudinal process. Second, the model only provides insights regarding how TVC observed values contribute to the heterogeneity in the longitudinal outcome; however, how a change in the TVC affects trajectory heterogeneity could be of more interest in practice. In this present work, we propose adding a decomposed TVC, based on the modeling framework proposed by \citet{Liu2022LCSM}, into the GMM to address the above two challenges. In the following sections, we describe how to add covariates into a GMM in detail and introduce the modeling framework that helps decompose a TVC.

\subsection*{Covaraites in Growth Mixture Models}\label{I:covariates}
As introduced above, adding covariates into a GMM has received lots of attention among researchers who utilize the SEM. In general, a GMM allows for the inclusion of TICs and TVCs. Existing studies have examined when, where, and how to add which TICs into a mixture model. As stated earlier, there are two types of TICs in the GMM: one type of TICs indicates group components, while the other type of TICs accounts for within-group differences in trajectories. In a GMM with the first-type TICs, the longitudinal outcome depends on a latent mixing component variable that, in turn, depends on the TICs. The inclusion of the first-type TICs enables one to understand how the likelihood of a trajectory belonging to an unobserved group changes conditional on the covariates. GMMs with the first-type TICs are popular among researchers employing the SEM. Earlier studies have demonstrated that adding this type of TICs can be achieved in a confirmatory approach through one-step models \citep{Bandeen1997one, Dayton1988one, Kamakura1994one, Yamaguchi2000one}, or more exploratory way through stepwise methods, such as two-step \citep{Bakk2017two, Liu2019BLSGMM} or three-step models \citep{Vermunt2010three, Bolck2004three, Asparouhov2014three}. One difference between the one-step and stepwise approaches is that the one-step method generates the estimates of the measurement parameters and the coefficients from the TICs to the latent classes simultaneously, while the stepwise method estimates them separately. Existing studies have shown that the one-step approach provides less biased and more accurate estimates, yet a `trial and error' fashion to construct a whole model and explore which TICs are needed to be included in a GMM requires lots of computational resources. More importantly, adding or removing a TIC may result in different solutions to the number of latent classes obtained from the Bayesian information criterion (BIC) based enumeration process, which has been well-defined and documented in the SEM literature \citep{Nylund2007number, Nylund2016number, Diallo2017number}.

Some recent works have proposed remedies to address such critiques of the one-step methods. For example, \citet{Kim2016expert, Hsiao2020mediation, Liu2021MoE, Liu2022PBLSGMM} have recommended constructing a one-step model in an exploratory fashion (also referred to as `the adjusted one-step approach'). Specifically, they suggested performing the enumeration processes without any TICs to have a stable solution for the number of latent classes and then constructing a whole model with the determined number of latent classes. Moreover, \citet{Liu2019BLSGMM, Liu2021MoE, Liu2022PBLSGMM} have demonstrated how to leverage machine learning techniques, such as feature extraction and feature selection in the SEM framework, to include the TICs that have impacts on trajectory heterogeneity in a GMM. In particular, \citet{Liu2019BLSGMM} proposed utilizing the exploratory factor analysis (EFA) \citep{Spearman1904factor} to transform candidate TICs from a large set to a manageable set with keeping the meaningful information of the original set. \citet{Liu2021MoE} recommended employing structural equation model forests (SEM forests) \citep{Brandmaier2016semForest} to identify the TICs that contributed the most to the trajectory heterogeneity. \citet{Liu2022PBLSGMM} further explored the two approaches and highlighted the difference in the insights obtained from these methods and the interpretations of such insights. 

Moreover, multiple existing studies, such as \citet{Asparouhov2014three, Kim2016expert, Hsiao2020mediation, Masyn2017expert, Liu2021MoE}, also examined mixture models with the second-type TICs. In the GMM with the second-type TICs, the outcome variable depends on both the latent component variable and the TICs: the growth factors of the longitudinal outcome are regressed on the TICs in each latent class, and the class-specific submodels are regressed on the latent component variable. Therefore, each submodel can be viewed as a multiple-indicator and multiple-cause (MIMIC) model \citep{Joreskog1975MIMIC, McArdle1987MIMIC}. Each MIMIC model has two components: (1) a measurement model where exogenous variables (i.e., repeated measurements of the outcome) indicate latent variables (i.e., growth factors) and (2) a structural model in which TICs are multiple-causal predictors of the latent variables. The inclusion of the second-type TICs allows for (1) identifying the heterogeneity in the TICs and (2) evaluating the heterogenous effects on the growth factors of the TICs across the latent classes of trajectories. GMMs with second-type TICs have received less attention than GMMs with first-type TICs. However, these earlier studies have demonstrated severe consequences of ignoring or misspecifying the second-type TICs in a mixture model by performing simulation studies, especially under the scenario that these TICs contribute to cluster separation. Such severe consequences may include misclassifying trajectories and/or generating biased estimates, 

\citet{Liu2021MoE} have demonstrated that the two types of TICs can simultaneously be included in a GMM. In addition, they showed that a covariate in a GMM is not necessarily a TIC by extending a longitudinal model proposed by \citet{Grimm2007multi} and regressing the development of mathematics ability on the repeated measurements of approach-to-learning in each latent class. The model successfully identified the heterogeneity in the variable approach-to-learning and the heterogeneous impacts on the mathematics development of the covariate approach-to-learning across students' latent classes. However, in the built model, the covariate approach-to-learning is not allowed to explain the variability of the growth factors of the development of mathematics ability. Moreover, the insights obtained from the model are limited to the effects of the absolute observations of the development of approach-to-learning. However, the impact of the change in the learning approach on intellectual development could be of more interest in practice. \citet[Chapter~8]{Grimm2016growth} also stated an inherent limitation of regressing a longitudinal outcome on a TVC in the SEM framework\footnote{According to \citet{Grimm2007multi}, a longitudinal model with a TVC could be constructed in the mixed-effects modeling framework and the SEM framework. However, the models built up in the SEM framework are able to provide more insights due to the flexibility of the framework. The SEM framework allows for estimating the time-varying impacts of a TVC and the covariances between the TVC and the growth factors of the longitudinal outcome.}: the full model has many parameters as there is no restricted structure on the TVC. Then the mean vector, variance-covariance matrix, and residuals of TVC are needed to be estimated. The estimation of covariances between repeated measures of TVC is challenging under some undesirable conditions. One solution to this is to fix the covariances to zero. However, it may not be valid if the TVC is expected to be somewhat stable. This limitation still exists when extending the model to the GMM framework. For example, in \citet{Liu2021MoE}, the covariances between the repeated measures of approach-to-learning are assumed to be zero in all latent classes, which may not be valid according to the developmental theory.

\subsection*{A Framework to Decompose Time-varying Covariates}\label{I:Decompose}
This section introduces a modeling framework to decompose a TVC and address the above three challenges. \citet{Liu2022LCSM} proposed a novel specification for latent change score modeling (LCSM) framework \citep{Zhang2012LCSM, Grimm2013LCSM, Grimm2013LCSM2} to fit longitudinal data. The new specification for the LCSM views the change that occurs in a time interval as the area under the curve (AUC) of the rate-of-change versus time ($r-t$) graph. The novel specification then allows for estimating the individual baseline value and interval-specific slopes. This feature provides a natural method to decompose a longitudinal variable into a trait feature (i.e., the baseline value) and a set of state features (i.e., interval-specific slopes). Note that both the baseline value and the interval-specific slopes are allowed to vary from individual to individual. Based on the interval-specific slopes, one is able to derive other types of state features, such as interval-specific changes and change-from-baseline values, which are also at the individual level. The novel specification and the two possible modifications enable us to model a TVC with a restricted structure, which in turn, reduces the number of parameters. 

The new specification for the LCSM is able to model longitudinal data with parametric or nonparametric functional forms \citep{Liu2022LCSM}. This article only focuses on the latter one, which can also be viewed as a latent basis growth model (LBGM) in the LCSM framework. We provide its path diagram with six repeated measurements in Figure \ref{fig:path_slp}. In the figure, $x_{j}$ and $x^{\ast}_{j}$ are the observed and latent true scores at $t=t_{j}$, respectively. In addition, $dx_{j+1}$ and $\gamma_{j}$ are the slope and relative rate in the $j^{th}$ time interval (i.e., from $t=t_{j}$ to $t=t_{j+1}$), respectively. Figure \ref{fig:path_slp} shows that there are two growth factors, $\eta^{[x]}_{0}$ and $\eta^{[x]}_{1}$, with this model specification, indicating the intercept and the slope in the first time interval (also referred to as the shape factor\footnote{In the SEM literature, the slope in the first time interval is also referred to as a shape factor, which can be scaled in multiple ways.} for a LBGM), respectively. With this specification, the relative rate during the first time interval (i.e., $\gamma_{1}$) is specified as $1$ so that $dx_{2}$ is the slope in this interval (i.e., $dx_{2}=\eta^{[x]}_{1}$). Similarly, the slope of each other time interval is defined as the product of the shape factor and the corresponding relative rate (i.e., $dx_{j+1}=\eta^{[x]}_{1}\times\gamma_{j}$ for $j\ge2$). As demonstrated in Figure \ref{fig:path_slp}, the latent true score at baseline is indicated by the growth factor $\eta^{[x]}_{0}$ (i.e., $x^{\ast}_{1}=\eta^{[x]}_{0}$). The latent true score at each post-baseline time point is defined as the sum of the latent true score at the previous time point and the product of an interval-specific slope and the corresponding interval length (i.e., $x^{\ast}_{j}=x^{\ast}_{j-1}+dx_{j}\times(t_{j}-t_{(j-1)})$). We then obtain observed scores by allowing a residual around the corresponding latent true score. Note that each time interval in Figure \ref{fig:path_slp} is in a diamond shape, suggesting a good feature of the specification: the LBGM is allowed to construct in the framework of individual measurement occasions by using `definition approaches' \citep{Mehta2000people, Mehta2005people, Sterba2014individually}. The `definition variables' are observed variables to adjust model coefficients to individual-specific values. For this specification, such individual-specific values are individual time intervals. As demonstrated in Figure \ref{fig:path_slp}, the model allows for separately estimating the initial status and the interval-specific slopes, which provides a natural way to decompose a TVC.

\figurehere{1}

In addition to the growth factors to indicate the initial status and the shape factor, a LBGM with the novel specification contains other latent variables, including interval-specific slopes (i.e., $dx_{j}$) and true scores (i.e., $x^{\ast}_{j}$) over time. The interval-specific slopes and true scores are derived from other parameters instead of being freely estimated. These non-estimable latent variables can be added into paths, serving as predictors in a model. With this idea, we are able to assess class-specific trait effects of a TVC by regressing the growth factors of a longitudinal outcome on the initial status in each class and evaluate class-specific state effects of a TVC by regressing each observed value of a longitudinal outcome on the corresponding interval-specific slope in each cluster. 

A slight modification by including additional latent variables on the model specified in Figure \ref{fig:path_slp} enables one to have a different set of state features. For example, we add $\delta x_{j}$ in Figure \ref{fig:path_chg} to define the change that occurs in the time interval from $t_{j-1}$ to $t_{j}$ (i.e., $\delta x_{j}=dx_{j}\times(t_{j}-t_{j-1})$). This modification allows for the decomposition of a TVC into the initial status and a collection of interval-specific changes. The explicit inclusion of interval-specific changes in a TVC enables them to serve as predictors of the observed values of a longitudinal outcome\footnote{Another possible modification is adding latent variables to represent the change-from-baseline values as \citet{Liu2022decompse}. This modification decomposes a TVC into the baseline value and a set of change-from-baseline values. Based on \citet{Liu2022decompse}, regressing the repeated measurements on the change-from-baseline values may result in biased estimates for a single group model. Therefore, we do not consider this modification in the present article.}. We consider these two ways to decompose a TVC. The estimated TVC initial status serves as the trait feature in both methods, whereas the interval-specific slopes or changes are two possible sets of state features. We then include a decomposed TVC in a GMM to evaluate the heterogeneity in the trait and state features and their heterogeneous trait and state effects on the longitudinal outcome across latent classes. The proposed models are built in the framework of individual measurement occasions to follow multiple existing studies that illustrate the `definition variables' approach in the SEM framework \citep{Sterba2014individually, Liu2019BLSGM, Liu2022LCSM} to avoid possible unacceptable solutions \citep{Blozis2008coding}. 

We organize the rest of this article as follows. We describe the model specification and model estimation of the GMMs with TICs and a decomposed TVC in the Method section. We then design a Monte Carlo simulation study to evaluate the proposed models. In particular, we examine the clustering and estimating effects of the proposed models. The clustering effects are evaluated by accuracy. The estimating effects are assessed by four performance metrics, including the relative bias, the empirical standard error (SE), the relative root-mean-squared-error (RMSE), and the empirical coverage probability (CP) for a nominal $95\%$ confidence interval. We perform a real-world analysis in the Application section to illustrate the proposed models. The Discussion section is crafted with practical considerations, methodological considerations, and future directions.  

\section*{Method}\label{Method}
\subsection*{Approach to Decomposing a Time-varying Covariate}\label{M:Decompose}
This section presents the statistical methods to decompose a TVC into a trait and a collection of state features introduced above. In particular, \citet{Liu2022LCSM} developed a novel specification for the nonparametric LCSM (i.e., the LBGM), which allows for estimating the initial status and interval-specific slopes and, therefore, provides a natural way to decompose a TVC into a trait and a set of state features. In the proposed specification, the LBGM with $J$ repeated measures is viewed as a linear piecewise function with $J-1$ segments following earlier studies \citep{McArdle2001LCM1} and \citep[Chapter~11]{Grimm2016growth}. The model is specified for the $i^{th}$ individual as
\begin{align}
&x_{ij}=x^{\ast}_{ij}+\epsilon^{[x]}_{ij},\label{eq:LBGM1}\\
&x^{\ast}_{ij}=\begin{cases}
\eta^{[x]}_{0i}, & (j=1) \\
x^{\ast}_{i(j-1)}+dx_{ij}\times(t_{ij}-t_{i(j-1)}), & (j=2, \dots, J)
\end{cases},\label{eq:LBGM2_1}\\
&dx_{ij}=\eta^{[x]}_{1i}\times\gamma_{j-1}\qquad (j=2, \dots, J). \label{eq:LBGM3}
\end{align}

The above three equations define a LBGM with the novel specification proposed by \citet{Liu2022LCSM}. Equation \ref{eq:LBGM1} writes an observed measurement (i.e., $x_{ij}$) as the sum of a latent true score (i.e., $x^{\ast}_{ij}$) and a residual (i.e., $\epsilon^{[x]}_{ij}$) for the $i^{th}$ individual at time $t_{j}$. Equation \ref{eq:LBGM2_1} defines the latent true scores at baseline and each post-baseline. Specifically, the true score is the growth factor indicating the intercept (i.e., $\eta^{[x]}_{0i}$) at baseline (i.e., $t=t_{i1}$). Meanwhile, at each post-baseline (i.e., $j\ge2$), the true score is expressed as a linear combination of the score at the prior time point $t_{i(j-1)}$ and the amount of true change from time $t_{i(j-1)}$ to $t_{ij}$. As suggested in Equation \ref{eq:LBGM2_1}, the interval-specific change is the product of the time interval (i.e., $t_{ij}-t_{i(j-1)}$) and the slope (i.e., $dx_{ij}$) in that interval. The interval-specific slope is further defined by the product of the slope of the first interval (i.e., the shape factor $\eta^{[x]}_{1i}$) and the corresponding relative rate (i.e., $\gamma_{j-1}$) as demonstrated in Equation \ref{eq:LBGM3}. Note that the subscript $i$ of time $t$ indicates that the recorded times can be individually different.

We can modify the model specified in Equation \ref{eq:LBGM2_1} slightly to explicitly include additional latent variables for the interval-specific changes, as Equations \ref{eq:LBGM2_21} and \ref{eq:LBGM2_22}
\begin{align}
&x^{\ast}_{ij}=\begin{cases}
\eta^{[x]}_{0i}, & (j=1)\\
x^{\ast}_{i(j-1)}+\delta x_{ij}, & (j=2, \dots, J)
\end{cases},\label{eq:LBGM2_21}\\
&\delta x_{ij}=dx_{ij}\times(t_{ij}-t_{i(j-1)})\qquad (j=2, \dots, J), \label{eq:LBGM2_22}
\end{align}
where $\delta x_{ij}$ indicates the amount of change that occurs between $t_{i(j-1)}$ and $t_{ij}$ of the $i^{th}$ individual. Therefore, Equations \ref{eq:LBGM1}, \ref{eq:LBGM2_21}, \ref{eq:LBGM2_22} and \ref{eq:LBGM3} together define a LBGM with the modified specification with explicit interval-specific changes (i.e., the model specified in Figure \ref{fig:path_chg}). The interval-specific slopes in Equation \ref{eq:LBGM3} or interval-specific changes in Equation \ref{eq:LBGM2_22} are allowed to be added to paths and serve as predictors in the SEM framework.

The above two specifications for the LBGM can be expressed in the same matrix form, which is only based on the freely estimable parameters in the model,
\begin{equation}\nonumber
\boldsymbol{x}_{i}=\boldsymbol{\Lambda}^{[x]}_{i}\times\boldsymbol{\eta}^{[x]}_{i}+\boldsymbol{\epsilon}^{[x]}_{i}
\end{equation}
where $\boldsymbol{x}_{i}$ is a $J\times1$ vector of the repeated measures of the TVC of individual $i$ (in which $J$ is the number of repeated measurements). In addition, $\boldsymbol{\eta}^{[x]}_{i}$ is a $2\times1$ vector of the growth factors of the TVC, representing the baseline value and the slope of the first time interval of the $i^{th}$ individual, respectively, while $\boldsymbol{\Lambda}^{[x]}_{i}$ is a $J\times2$ matrix of the corresponding factor loadings, 
\begin{equation}\nonumber
\boldsymbol{\Lambda}^{[x]}_{i}=\begin{pmatrix}
1 & 0 \\
1 & \gamma_{1}\times(t_{i2}-t_{i1}) \\
1 & \sum_{j=2}^{3}\gamma_{j-1}\times(t_{ij}-t_{i(j-1)}) \\
\dots & \dots \\
1 & \sum_{j=2}^{J}\gamma_{j-1}\times(t_{ij}-t_{i(j-1)}) \\
\end{pmatrix},
\end{equation}
of which all elements in the first column are $1$ since they are the factor loadings of the initial status of the TVC. The $j^{th}$ element of the second column is the cumulative value of the relative rate (i.e., $\gamma_{j}$) over time to $t_{ij}$, so its product with $\eta^{[x]}_{1i}$ represents the amount of change-from-baseline at occasion $t_{j}$. Additionally, $\boldsymbol{\epsilon}^{[x]}_{i}$ is a $J\times1$ vector of residuals of the TVC for the $i^{th}$ individual. The growth factors $\boldsymbol{\eta}^{[x]}_{i}$ can be further written as
\begin{equation}\nonumber
\boldsymbol{\eta}^{[x]}_{i}=\boldsymbol{\mu}^{[x]}_{\boldsymbol{\eta}}+\boldsymbol{\zeta}^{[x]}_{i},
\end{equation}
where $\boldsymbol{\mu}^{[x]}_{\boldsymbol{\eta}}$ is the mean vector of the TVC growth factors, while $\boldsymbol{\zeta}^{[x]}_{i}$ is the vector of deviations of individual $i$ from the corresponding growth factor means. More technical details are documented in \citet{Liu2022LCSM}.

\subsection*{Model Specification of Growth Mixture Model with Time-invariant Covariates and A Decomposed Time-varying Covariate}\label{M:Specification}
This section presents the model specification for the proposed GMMs with TICs and a decomposed TVC. In this article, we build a GMM with two first-type TICs, one second-type TICs, and a decomposed TVC for illustration purposes. Specifically, the latent mixing component variable of the GMM depends on the two first-type TICs, which divides the sample into multiple latent classes. In each latent class, we regress the growth factors of the longitudinal outcome on the second-type TIC and the trait feature of the TVC and regress each post-baseline value of the longitudinal outcome on the corresponding state feature. The longitudinal outcome may take any function in a GMM; therefore, we only provide a general model specification instead of pre-specifying them in the equations. Suppose that we have $K$ latent classes in total. The proposed GMM can be expressed as
\begin{align}
&p(\boldsymbol{y}_{i}|z_{i}=k,\boldsymbol{x}_{gi}, {x}_{ei}, \boldsymbol{x}_{i})=\sum_{k=1}^{K}\pi(z_{i}=k|\boldsymbol{x}_{gi})\times p(\boldsymbol{y}_{i}|z_{i}=k,{x}_{ei}, \boldsymbol{x}_{i}),\label{eq:MoE}\\
&\pi(z_{i}=k|\boldsymbol{x}_{gi})=\begin{cases}
\frac{1}{1+\sum_{k=2}^{K}\exp(\beta_{g0}^{(k)}+\boldsymbol{\beta}_{g}^{(k)T}\boldsymbol{x}_{gi})}& \text{reference Group ($k=1$)}\\
\frac{\exp(\beta_{g0}^{(k)}+\boldsymbol{\beta}_{g}^{(k)T}\boldsymbol{x}_{gi})} {1+\sum_{k=2}^{K}\exp(\beta_{g0}^{(k)}+\boldsymbol{\beta}_{g}^{(k)T}\boldsymbol{x}_{gi})} & \text{other Groups ($k=2,\dots, K$)}
\end{cases},\label{eq:Gating}\\
&\begin{pmatrix}\boldsymbol{x}_{i} \\ \boldsymbol{y}_{i}
\end{pmatrix}|(z_{i}=k)=\begin{pmatrix}
\boldsymbol{\Lambda}^{[x]}_{i} & \boldsymbol{0} \\
\boldsymbol{0} & \boldsymbol{\Lambda}^{[y]}_{i}
\end{pmatrix}\times\begin{pmatrix} \boldsymbol{\eta}^{[x]}_{i} \\ \boldsymbol{\eta}^{[y]}_{i}
\end{pmatrix}|(z_{i}=k)+\kappa^{(k)}_{1}\times\begin{pmatrix} \boldsymbol{0} \\ \boldsymbol{dx_{i}}
\end{pmatrix}|(z_{i}=k)+\begin{pmatrix}
\boldsymbol{\epsilon}^{[x]}_{i} \\ \boldsymbol{\epsilon}^{[y]}_{i}
\end{pmatrix}|(z_{i}=k), \label{eq:Experts1}\\
&\boldsymbol{\eta}^{[y]}_{i}|(z_{i}=k)=\boldsymbol{\alpha}^{(k)[y]}+\begin{pmatrix}\boldsymbol{\beta}^{(k)}_{\text{TIC}} & \boldsymbol{\beta}^{(k)}_{\text{TVC}}\end{pmatrix}\times\begin{pmatrix}x_{ei} \\ \eta^{[x]}_{0i}\end{pmatrix}|(z_{i}=k)+\boldsymbol{\zeta}^{[y]}_{i}|(z_{i}=k), \label{eq:Experts2}\\
&\boldsymbol{\eta}^{[x]}_{i}|(z_{i}=k)=\boldsymbol{\mu}^{(k)[x]}_{\boldsymbol{\eta}}+\boldsymbol{\zeta}^{[x]}_{i}|(z_{i}=k), \label{eq:Experts3}
\end{align}
for the $i^{th}$ individual in the $k^{th}$ latent class. 

Equation \ref{eq:MoE} defines a GMM that combines $K$ submodels. Following \citet{Liu2021MoE}, we define $\boldsymbol{x}_{gi}$ and $x_{ei}$ as the first-type TICs and the second-type TIC, respectively. Note that $\boldsymbol{x}_{gi}$ is a vector while $x_{ei}$ is a scalar in the equations since we specify two first-type TICs and only one second-type TIC in this illustration. In addition, $\boldsymbol{x}_{i}$ is the vector of the repeated measures of the TVC as defined in the previous section, and $z_{i}$ is the membership of the $i^{th}$ individual. As indicated in Equation \ref{eq:MoE}, the latent mixing component variable $\pi()$ depends on the first-type TICs and divides the sample into $K$ latent classes. In each class, the longitudinal outcome depends on the second-type TIC and the TVC. Note that there are two constraints on the mixing component variable: (1) $0\le\pi(z_{i}=k|\boldsymbol{x}_{gi})\le1$ and (2) $\sum_{k=1}^{K}\pi(z_{i}=k|\boldsymbol{x}_{gi})=1$. Equation \ref{eq:Gating} defines the mixing component variable as logistic functions of the first-type TICs, where $\beta^{(k)}_{0}$ and $\boldsymbol{\beta}^{(k)}$ are the intercept and coefficients of the logistic functions, respectively.

Equations \ref{eq:Experts1}, \ref{eq:Experts2} and \ref{eq:Experts3} together define the submodel (i.e., the within-class model). In Equation \ref{eq:Experts1}, the longitudinal outcome $\boldsymbol{y}_{i}$ is a $J\times1$ vector (where $J$ is the number of measurement occasions) similar to the TVC $\boldsymbol{x}_{i}$. Moreover, $\boldsymbol{\eta}^{[y]}_{i}$ is a $C\times1$ vector of growth factors (where $C$ is the number of growth factors of the longitudinal outcome), and $\boldsymbol{\Lambda}^{[y]}_{i}$ is a $J\times C$ matrix of the corresponding factor loadings. Note that the subscript $i$ in the matrices of factor loadings indicates that the GMM is built up in the framework of individual measurement occasions. Additionally, $\boldsymbol{\epsilon}^{[y]}_{i}$ is a $J\times1$ vector of residuals of the longitudinal outcome for individual $i$. In Equation \ref{eq:Experts1}, $\boldsymbol{dx_{i}}$ is a $J\times1$ vector of interval-specific slopes of the TVC. We can further express the interval-specific slopes as $\boldsymbol{dx_{i}}=\begin{pmatrix}0 & dx_{i2} & dx_{i3} & \dots & dx_{iJ}\end{pmatrix}$, where the first element is $0$, while $dx_{ij}$ is the slope in time interval $j-1$ of individual $i$. Accordingly, $\kappa^{(k)}_{1}$ is the class-specific state effect of the TVC, characterizing how the measure of the longitudinal outcome at $t_{j}$ is influenced by the slope in the previous time interval (i.e., between $t_{j-1}$ and $t_{j}$) in the $k^{th}$ latent class. In Equation \ref{eq:Experts1}, $\boldsymbol{0}$ is also a $J\times1$ vector.

Equation \ref{eq:Experts2} further regresses the growth factors of the longitudinal outcome on the second-type TIC and the growth factor that indicates the initial status of the TVC. In Equation \ref{eq:Experts2}, $\boldsymbol{\alpha}^{(k)[y]}$ is a $C\times1$ vector of class-specific intercepts of the growth factors, $\boldsymbol{\beta}^{(k)}_{\text{TIC}}$ ($\boldsymbol{\beta}^{(k)}_{\text{TVC}}$) is $C\times1$ vector of class-specific regression coefficients from the second-type TIC (initial status of the TVC) to the growth factors. In addition, $x_{ei}$ is the value of the second-type TIC and $\eta^{[x]}_{0i}$ is the TVC initial status of the $i^{th}$ individual, and $\boldsymbol{\zeta}^{[y]}_{i}$ is a $C\times1$ vector of deviations of the individual $i$ from the conditional means of growth factors. Equation \ref{eq:Experts3} further defines the growth factors of the TVC in each latent class as the sum of the class-specific mean vector of the growth factors and the vector of deviations of the $i^{th}$ individual from the corresponding growth factor means. 

Similarly, the submodel of the proposed GMM with TICs and a TVC with the other type of decomposition for individual $i$ can be expressed as
\begin{equation}\label{eq:Experts1_2}
\begin{pmatrix}\boldsymbol{x}_{i} \\ \boldsymbol{y}_{i}
\end{pmatrix}|(z_{i}=k)=\begin{pmatrix}
\boldsymbol{\Lambda}^{[x]}_{i} & \boldsymbol{0} \\
\boldsymbol{0} & \boldsymbol{\Lambda}^{[y]}_{i}
\end{pmatrix}\times\begin{pmatrix} \boldsymbol{\eta}^{[x]}_{i} \\ \boldsymbol{\eta}^{[y]}_{i}
\end{pmatrix}|(z_{i}=k)+\kappa^{(k)}_{2}\times\begin{pmatrix} \boldsymbol{0} \\ \boldsymbol{\delta x_{i}}
\end{pmatrix}|(z_{i}=k)+\begin{pmatrix}
\boldsymbol{\epsilon}^{[x]}_{i} \\ \boldsymbol{\epsilon}^{[y]}_{i}
\end{pmatrix}|(z_{i}=k),
\end{equation}
where $\boldsymbol{\delta x_{i}}$ is a $J\times1$ vector of interval-specific changes of TVC. We further express interval-specific changes as $\boldsymbol{\delta x_{i}}=\begin{pmatrix}0 & \delta x_{i2} & \delta x_{i3} & \dots & \delta x_{iJ}\end{pmatrix}$. Similar to $\boldsymbol{dx_{i}}$, the first element of $\boldsymbol{\delta x_{i}}$ is $0$, and $\delta x_{ij}$ is the change that occurs in the $(j-1)^{th}$ time interval for individual $i$. Therefore, $\kappa^{(k)}_{2}$ is a class-specific state effect of the TVC, depicting how the value of the longitudinal outcome at $t_{j}$ is affected by the amount of change in the previous time interval in cluster $k$. So Equations \ref{eq:MoE}, \ref{eq:Gating}, \ref{eq:Experts1_2}, \ref{eq:Experts2} and \ref{eq:Experts3} together define the GMM with TICs and a decomposed TVC into the initial status and interval-specific changes.

In GMMs, the longitudinal outcome $\boldsymbol{y}_{i}$ may take any functional form, either a linear or nonlinear function, and the pre-specified functions are allowed to vary from class to class. The specifications of functional forms depend on the change patterns demonstrated by raw data and the research questions of particular interest. In Table \ref{tbl:traj_summary}, we list functional forms, growth factors and corresponding interpretations, and factor loadings for the linear growth curve and four commonly used nonlinear trajectories, namely the quadratic, negative exponential, Jenss-Bayley, and linear-linear function with an unknown fixed knot.  

\tablehere{1}

\subsection*{Model Estimation}\label{M:Estimation}
This section describes how to estimate the proposed GMMs with TICs and a decomposed TVC. We make the following three assumptions to simplify model estimation. First, the second-type TIC and the growth factors of the TVC in each latent class are normally distributed; that is, $x_{ei}|(z_{i}=k)\sim N(\mu^{[k]}_{x}, \phi^{[k]}_{x})$ and $\boldsymbol{\zeta}^{[x]}_{i}|(z_{i}=k)\sim \text{MVN}(\boldsymbol{0}, \boldsymbol{\Phi}^{(k)[x]}_{\boldsymbol{\eta}})$, in which $\mu^{[k]}_{x}$ and $\phi^{[k]}_{x}$ are the class-specific mean and variance of the second-type TIC, while $\boldsymbol{\Phi}^{(k)[x]}_{\boldsymbol{\eta}}$ is a $2\times2$ class-specific variance-covariance matrix of the TVC growth factors. Second, we assume that the growth factors of the longitudinal outcome are normally distributed conditional on the second-type TIC and the true score of the TVC initial status in each latent class. Therefore, we have $\boldsymbol{\zeta}^{[y]}_{i}|(z_{i}=k)\sim \text{MVN}(\boldsymbol{0}, \boldsymbol{\Psi}^{(k)[y]}_{\boldsymbol{\eta}})$ in which $\boldsymbol{\Psi}^{(k)[y]}_{\boldsymbol{\eta}}$ is a $C\times C$ class-specific unexplained variance-covariance matrix of the growth factors of the longitudinal outcome. The third assumption is that the variable-specific residuals are identical and independent normal distributions and that the residual correlations are homogeneous over time, then we have 
\begin{equation}\nonumber
\begin{pmatrix}
\boldsymbol{\epsilon}^{[x]}_{i} \\ \boldsymbol{\epsilon}^{[y]}_{i}\end{pmatrix}|(z_{i}=k)\sim \text{MVN}\bigg(\begin{pmatrix}
\boldsymbol{0} \\ \boldsymbol{0}\end{pmatrix}, \begin{pmatrix} \theta^{(k)[x]}_{\epsilon}\boldsymbol{I} &
\theta^{(k)[xy]}_{\epsilon}\boldsymbol{I} \\ & \theta^{(k)[y]}_{\epsilon}\boldsymbol{I} \end{pmatrix}\bigg),
\end{equation}
where $\boldsymbol{I}$ is a $J\times J$ identify matrix, $\theta^{(k)[x]}$ and $\theta^{(k)[y]}$ are the class-specific residual variance of the TVC and the longitudinal outcome, respectively, and $\theta^{(k)[xy]}$ is the class-specific residual covariance. We provide the expressions of the model-implied mean vector and variance-covariance structure of the bivariate longitudinal variables in each latent class in Appendix \ref{supp1}.

The parameters in the proposed GMMs with TICs and a decomposed TVC include the class-specific mean ($\mu^{(k)}_{x}$) and variance ($\phi^{(k)}_{x}$) of the second-type TIC, the class-specific mean vector ($\boldsymbol{\mu}^{(k)[x]}_{\boldsymbol{\eta}}$) and variance-covariance matrix ($\boldsymbol{\Phi}^{(k)[x]}_{\boldsymbol{\eta}}$) of the TVC growth factors, the class-specific intercepts ($\boldsymbol{\alpha}^{(k)[y]}$) and unexplained variance-covariance ($\boldsymbol{\Psi}^{(k)[y]}_{\boldsymbol{\eta}}$) of the growth factors of the longitudinal outcome, the class-specific effects on the growth factors of the longitudinal outcome of the TIC ($\boldsymbol{\beta}^{(k)}_{\text{TIC}}$), the class-specific trait effects ($\boldsymbol{\beta}^{(k)}_{\text{TVC}}$), the class-specific state effect of the TVC ($\kappa^{(k)}_{1}$ or $\kappa^{(k)}_{2}$ for the two types of decomposition, respectively), and the class-specific residual variances ($\theta^{(k)[x]}$ and $\theta^{(k)[y]}$) and covariance ($\theta^{(k)[xy]}$). The relationship between the second-type TIC and TVC is captured by the class-specific correlation ($\rho^{(k)}_{\text{BL}}$) between the TIC and the trait feature of the TVC (i.e., the true score of the TVC initial status). In addition, we also need to estimate the coefficients in the logistic regressions. We define $\boldsymbol{\Theta}$ as
\begin{align}
\boldsymbol{\Theta}=&\{\mu^{(k)}_{x}, \phi^{(k)}_{x}, \boldsymbol{\mu}^{(k)[x]}_{\boldsymbol{\eta}}, \boldsymbol{\Phi}^{(k)[x]}_{\boldsymbol{\eta}}, \gamma^{(k)}_{2}, \dots, \gamma^{(k)}_{J-1}, \boldsymbol{\alpha}^{(k)[y]}, \boldsymbol{\Psi}^{(k)[y]}_{\boldsymbol{\eta}}, \nonumber\\
&\boldsymbol{\beta}^{(k)}_{\text{TIC}}, \boldsymbol{\beta}^{(k)}_{\text{TVC}}, \kappa^{(k)}_{1/2}, \rho^{(k)}_{\text{BL}}, \theta^{(k)[x]}, \theta^{(k)[y]}, \theta^{(k)[xy]}, \beta^{(k)}_{0}, \boldsymbol{\beta}^{(k)}\} \nonumber\\
&k=2,\dots,K \text{ for } \beta^{(k)}_{0}, \boldsymbol{\beta}^{(k)} \nonumber\\
&k=1,\dots,K \text{ for other parameters}\nonumber
\end{align}
to list the parameters in the specified GMMs. In addition, as demonstrated in Table \ref{tbl:traj_summary}, there may exist additional class-specific growth coefficients to capture the change patterns of the longitudinal outcome, such as class-specific coefficients $b^{(k)}$ in the negative exponential growth curve, $c^{(k)}$ in the Jenss-Bayley trajectory, and $\gamma^{(k)}$ in the bilinear spline functional form. These additional growth coefficients also need to be estimated in the proposed model.

We utilize the full information maximum likelihood (FIML) technique in the article to estimate the proposed GMMs with TICs and a decomposed TVC. The FIML approach allows for the heterogeneity of individual contributions to the likelihood. The proposed GMMs are built using the R package \textit{OpenMx} with the optimizer CSOLNP \citep{OpenMx2016package, Pritikin2015OpenMx, Hunter2018OpenMx, User2020OpenMx}. We provide \textit{OpenMx} code on the GitHub website (\url{https://github.com/xxxx}) (computing code will be uploaded upon acceptance) to demonstrate how to apply the proposed models. The proposed GMMs can also fit with other SEM software such as \textit{Mplus 8}; the corresponding code is also provided on the GitHub website for researchers interested in using it.

\section*{Model Evaluation}\label{Evaluation}
We evaluate the proposed GMMs with TICs and a decomposed TVC using a simulation study with two goals. The first goal is to assess the performance metrics and clustering effect of the proposed GMMs. The performance measures include relative bias, empirical SE, RMSE, and empirical coverage for the $95\%$ CI. The definitions and estimators of these four measures are provided in Table \ref{tbl:metric}. The clustering effect is characterized by accuracy, defined as the proportion of all correctly classified instances \citep[Chapter~1]{Bishop2006pattern} as true memberships are available in the simulation study. With equation
\begin{equation}\nonumber
p(z_{i}=k)=\frac{g(z_{i}=k|\boldsymbol{x}_{gi})p(\boldsymbol{y}_{i}|z_{i}=k, {x}_{ei}, \boldsymbol{x}_{i})}{\sum_{k=1}^{K}g(z_{i}=k|\boldsymbol{x}_{gi})p(\boldsymbol{y}_{i}|z_{i}=k, {x}_{ei}, \boldsymbol{x}_{i})},
\end{equation}
we are able to estimate membership for each individual. The second goal is to compare the proposed GMMs with the corresponding full mixture model (i.e., GMMs with all TICs but only with the trait feature of the TVC) and finite mixture model (i.e., GMM without any covariates)\footnote{Note that the finite mixture model can be viewed as a reduced model of the proposed GMM by restricting all coefficients from covariates as $0$. However, the full mixture model, in which the growth factors are regressed on the observed initial status in each latent class, is not a reduced model of the proposed GMM, where the growth factors are regressed on the true initial status in each cluster. }. The comparison is based on Dumenci's Latent Kappa Statistics \citep{Dumenci2011kappa, Dumenci2019knee}, performance metrics, and clustering effect. 

\tablehere{2}

Similar to the Method section, we consider two first-type TICs, one second-type TIC, and a decomposed TVC in the simulation design. Moreover, we assume the longitudinal outcome takes the linear-linear functional form with an unknown fixed knot defined in Table \ref{tbl:traj_summary} for three considerations. First, in each submodel, the variance of longitudinal outcome is divided into three parts, the variance explained by growth factors, the variance explained by state features of a TVC, and an unexplained variance (i.e., the residual). Therefore, including state features of the TVC in a GMM may affect the estimation of the growth factors of the longitudinal outcome. In practice, the majority of longitudinal processes are nonlinear. One advantage of the linear-linear piecewise function over other parametric trajectories is that it allows for assessing nonlinear change patterns with local linear functions. Therefore, this functional form is capable of explicitly evaluating the effects on the estimation of the slope of the TVC in each linear stage, which is unrealizable by other parametric nonlinear functions where each measurement is a composite of multiple growth factors. Second, \citet{Liu2021MoE} have documented the simulation results for GMMs with the two types of TICs, which allows us to only focus on the simulation conditions related to the trait and state effects of the TVC, and in turn, reduces the size of simulation conditions in this project. Third, it is feasible to extend the simulation result observed from the GMMs with the linear-linear functions to other functional forms.

The number of repetitions in this simulation design is set as $S=1,000$, determined by an empirical method proposed by \citet{Morris2019simulation}. In particular, we conducted a pilot simulation and observed that the standard errors of all coefficients, except those related to the initial status of longitudinal variables, were less than $0.15$. This suggests that at least $900$ repetitions are needed to keep the Monte Carlo standard error of the bias less than $0.005$\footnote{According to \citet{Morris2019simulation}, the equation of the Monte Carlo standard error of bias can be expressed as $\text{Monte Carlo SE(Bias)}=\sqrt{Var(\hat{\theta})/S}$. }. We then decided to perform a simulation study with $1,000$ replications to be more conservative. 

\subsection*{Design of Simulation Study}
Table \ref{tbl:simu_design} lists all conditions we considered for the proposed GMMs with TICs and a decomposed TVC in the simulation design. As stated earlier, we fixed the conditions examined in existing studies to limit the size of the simulation conditions in this research project. For example, we selected ten scaled and equally spaced study waves since \citet{Liu2022decompse} have shown that the one-group longitudinal model with linear-linear functional form and a TVC performed decently regarding the four performance metrics and that fewer study waves (e.g., six) of the longitudinal outcome did not affect model performance meaningfully. Following earlier studies, we obtained individual measurement occasions by allowing for a moderate time window ($-0.25$, $0.25$) around each study wave \citep{Coulombe2015ignoring}. In addition, two allocation ratios were considered in the simulation design, 1:1 and 1:2, which are roughly controlled by the intercept coefficients of the logistic functions. The class mixing proportions 1:1 was selected as it is a balanced allocation, while the other level of the allocation ratio helps examine model performance in an unbalanced condition (that is presumably more challenging) and understand how the sample size (of submodels) affects the proposed GMMs. 

\tablehere{3}

In the simulation design, we manipulated the conditions related to the TVC, cluster separation, and allocation ratio, which presumably affect clustering algorithms and TVC effects. A critical characteristic of a model-based clustering model is its ability to detect sample heterogeneity and estimate parameters of interest. Intuitively, the algorithm should perform better under conditions with greater separation between unobserved groups. The distance between latent classes of a mixture model is captured by the difference between the class-specific density functions, which is mainly gauged by two measures, the Mahalanobis distance between class-specific growth factors of the longitudinal outcome and the difference in the knot locations \citep{Kohli2015PLGC, Liu2019BLSGMM, Liu2021MoE}. In this project, we fix class-specific mean vectors and variance-covariance matrices of growth factors of the longitudinal outcome to keep the Mahalanobis distance as $0.86$, a small distance according to \citet{Kohli2015PLGC}. We set $1.0$, $1.5$, and $2.0$ as small, medium, and large differences between knot locations, respectively \citep{Liu2019BLSGMM, Liu2021MoE}. As pointed out by \citet{Liu2021MoE}, the covariates that explain within-class variability may also affect the distance between two classes, especially when the distributions of covariates are different across latent classes. Therefore, we set the joint distributions of the second-type TIC and the TVC baseline value to be the same across latent classes in the simulation study to minimize their possible impact on the distance between clusters. Moreover, we considered three scenarios of interval-specific slopes. In the first scenario, we set the same state features across classes. In the other two scenarios, the state features were set differently, either by different shape factors or different relative rates. In addition, the trait effects and state effects were set differently in the two latent classes. In addition, we considered two levels of the residual variance of the longitudinal outcome.

\subsection*{Data Generation of Simulation Step}
We utilized a two-step data generation for each condition of the two GMMs listed in Table \ref{tbl:simu_design}. First, we generated the first-type TICs, based on which we obtained the membership $z_{i}$ for each individual. Second, we simultaneously generate the longitudinal outcome, the second-type TIC, and the TVC for each cluster. The general steps of the simulation study are: 
\begin{enumerate}
\item{Obtain membership $z_{i}$ for the $i^{th}$ individual:}
\begin{enumerate}
\item{Generate data matrix of the first-type TICs $\boldsymbol{x}_{g}$,}
\item{Calculate the probability vector for each individual based on the first-type TICs and a collection of pre-specified logistic coefficients with a logic link, and assign each entry to the component with the highest probability,}
\end{enumerate}
\item {Generate the second-type TIC, the growth factors of the TVC, and the growth factors of the longitudinal outcome for each latent class using the R package \textit{MASS} \citep{Venables2002Statistics},}
\item {Generate a time structure with $J=10$ waves $t_{j}$ and obtain individual measurement occasions by allowing for a time window around each wave $t_{ij}\sim U(t_{j}-\Delta, t_{j}+\Delta)$ ($\Delta=0.25$), }
\item {Calculate factor loadings for the TVC and the longitudinal outcome for each latent class, which are functions of the individual measurement occasions and additional class-specific growth coefficient(s) (i.e., the relative rates of the TVC or the unknown knot for the longitudinal outcome),}
\item {Calculate a collection of state features of the TVC for each latent class, which are the interval-specific slopes or changes,}
\item {Calculate the true scores of repeated measurements for the TVC and the longitudinal outcome: the former is based on the TVC growth factors and the corresponding factor loadings, while the latter is based on the longitudinal outcome growth factors, factor loadings, and the interval-specific slopes or changes of the TVC, then add the residual matrix to the true scores of  the TVC and the longitudinal outcome,}
\item {Implement the proposed GMM with each decomposition method, estimate the parameters, and construct the corresponding $95\%$ Wald confidence intervals,}
\item {Repeat the above steps until achieving $1,000$ convergent solutions.}
\end{enumerate}

\section*{Results}
\subsection*{Model Convergence}\label{r:converge}
This section summarizes the convergence rate\footnote{In this present article, convergence is defined as the achievement of \textit{OpenMx} status code $0$, indicating a successful optimization with up to $10$ attempts with different sets of starting values.} of the proposed GMMs with two types of TICs and a decomposed TVC. In general, the proposed GMMs converged well across all conditions. Specifically, out of $36$ conditions for the GMM with a decomposed TVC into interval-specific slopes, $12$ conditions reported $100\%$ of convergence rate. It is noticed that all of these conditions with $100\%$ of convergence rate were those with a large difference in knot locations (i.e., the difference was $2$). In addition, $20$ of the rest conditions reported a convergence rate above $97\%$. The worst scenario regarding the non-convergence rate is $37/1037$, implying that the simulation process described above needs to be repeated $1037$ times to generate $1,000$ replications with a convergent solution. It occurred under the condition of balanced allocation, the small difference between knot locations, the TVC having different shape factors but the same relative rates across clusters, and the small residual variance. The convergence rates of the GMM with a decomposed TVC into interval-specific changes were similar. We only kept the converged replications for further evaluation. 

\subsection*{Performance Measures}\label{R:Est}
This section recapitulates the four performance metrics, including the relative bias, empirical SE, relative RMSE, and CP for each parameter of interest of the proposed GMMs. Given the number of parameters and the size of simulation conditions, we first calculated each performance measure across $1,000$ replications for each parameter of interest under each condition. Then we summarized the values of each performance measure for each parameter of interest across all conditions as the corresponding median and range. Based on the result of our simulation study, the magnitudes of the performance metrics of each parameter between the two models were similar. Specifically, the magnitude of relative biases and relative RMSE across conditions and parameters of interest was less than $10\%$ and $0.5$, respectively. The CPs of all parameters of interest except for the knots were around $95\%$. These results suggest that both GMMs are capable of generating unbiased and accurate point estimates with target CPs in general. The summary statistics of the four performance metrics of the two proposed GMMs are provided in Tables S1 and S2 in the Online Supplementary Document. The CPs of class-specified knots might be unsatisfactory. To examine the patterns, we provided the plots of CPs of cluster-specific knots, stratified by the cluster separation, in Figures S1 and S2 in the Online Supplementary Document. The CPs, which might be conservative under some challenging conditions, were still around $95\%$ under the conditions with the large separation. This finding aligns with results of simulation studies in earlier research works such as \citet{Liu2021MoE}.

\subsection*{Clustering Effects}\label{R:Clus}
In this section, we evaluate the clustering effect of the proposed GMMs, gauged by mean accuracy values across $1000$ replications for the conditions listed in Table \ref{tbl:simu_design}. We calculated the mean accuracy value across $1000$ replications for each model under each condition. In the Online Supplementary Document, we provided the plots of the mean accuracy values, stratified by cluster separation and differences in TVCs of two latent classes, in Figures S3 and S4. The proposed GMMs performed well in separating trajectories into latent classes since the accuracy value achieved at least $88\%$ across all conditions. The mean accuracy values were the highest under the conditions with the large cluster separation, followed by those with the medium separation and then the small separation. In addition, allowing different state features across latent classes, either by different shape factors or different relative rates, improved the accuracy values, as shown in Figures S3 and S4. It is within our expectations to see. The difference in state features between clusters increases the cluster separation; the larger the cluster separation is, the easier the clustering algorithm can tell them apart, leading to greater accuracy values. 

\subsection*{Comparison Among Models}\label{R:compare}
This section compares the proposed GMMs with a decomposed TVC to the corresponding full and finite mixture models. We first evaluated how the exclusion of the decomposed TVC or its state features affects the clustering membership captured by Dumenci's Latent Kappa statistics. We first calculated the mean value of Dumenci's Latent Kappa statistics across $1000$ repetitions. Then we plotted these values stratified by the cluster separation and the difference in TVC state features between clusters in Figures S5 and S6 for each condition of each GMM with a TVC, respectively. We first noticed that the mean values of latent Kappa between the proposed models and the corresponding full mixture models under the conditions with the same TVC state features were above $0.8$, indicating almost perfect agreement \citep{Landis1977kappa, Nakazawa2019fmsb}. It suggests that excluding the TVC state features from a GMM did not affect the membership meaningfully under such conditions. On the contrary, removing TVC state features from a GMM affected the membership to some extent when the state features differed across clusters, and the corresponding mean latent Kappa was less than $0.70$. Excluding TICs from a GMM may further affect the clustering membership slightly. 

In addition, the performance measures and accuracy of the full and finite mixture models were worse than the corresponding proposed model. Some estimates from the models without a TVC exhibited some bias greater than $10\%$. For example, the magnitude of relative bias of the effects of the second-type TICs may achieve $50\%$. One possible explanation for such unsatisfactory performance of the full mixture model is that we regressed the growth factors of the longitudinal outcome on the observed TVC initial status rather than the corresponding true scores as we did in the proposed GMMs. The four performance metrics of the model without a TVC were also provided in the Online Supplementary Materials (Tables S3-S6). We also plotted the mean accuracy values of the reduced mixture models in Figures S3 and S4. From the figures, we observed that the mean accuracy values of the mixture model without a TVC were smaller than those with a TVC, suggesting less satisfactory clustering effects of the models excluding a TVC. 

\section*{Application}
This section demonstrates how to employ the proposed GMMs with the two types of TICs and a decomposed TVC to perform a real-world analysis with two goals. The first goal is to explore how the inclusion of a TVC when modeling a mixture model affects the estimated membership and growth factors of the longitudinal outcome. Based on this, we aim to provide a feasible set of recommendations for applying the proposed models. For this application, a random sample of 500 students was selected from the Early Childhood Longitudinal Study, Kindergarten Cohort: 2010-2011 (ECLS-K: 2011). The extracted dataset contains non-missing records of repeated reading and mathematics item response theory (IRT) scores with baseline family income, parents' education, and teacher-reported inhibitory control\footnote{There are $n=18174$ participants in ECLS-K: 2011. After removing missing values (i.e., records with any NaN/-9/-8/-7/-1) in the selected variables, $2140$ students were held.}.

ECLS-K: 2011, which starts from the 2010-2011 school year, is a longitudinal study to gather information on child development from approximately $900$ kindergarten programs across the United States. Students' reading and mathematics skills were evaluated in nine waves, each semester in the first three years (i.e., from Grade K to Grade 2) and then only the spring semester in the rest of three years (i.e., from Grade 3 to Grade 5). Only about one-third of students were assessed in the fall semesters of Grades 1 and 2 \citep{Le2011ECLS}. This survey has two types of time metrics: study waves, indicated by school semesters, and students' real age (in months). We used students' real age and converted it to age-in-years for individual measurement occasions. In this application, we viewed mathematics development as the longitudinal outcome while reading development as the TVC. In addition, the baseline socioeconomic status, including the highest education among parents and family income, and the baseline teacher-reported inhibitory control as the first and second types of covariates, respectively.

\subsection*{Enumeration Process}
This section demonstrated how to determine the number of latent classes for the proposed GMMs with the two types of TICs and a decomposed TVC. Following the SEM literature convention, we conducted the enumeration process without covariates. We first fit one-, two-, three- and four-class models for mathematics development. All four models converged, and the corresponding estimated likelihood values, information criteria, including Akaike information criterion (AIC) and BIC, and class-specific residual variances are provided in Table \ref{tbl:info}. From the table, we noticed that the model with three latent classes has the smallest BIC, leading to the conclusion that the optimal number of latent classes for mathematics development was three from the statistical perspective. The optimal model suggested the estimated mixing proportions of the student clusters are $20.20\%$, $42.20\%$, and $37.60\%$, respectively.

\tablehere{4}

\subsection*{Proposed Models}
We then fit the proposed GMMs with the two types of TICs and a decomposed TVC to evaluate how the baseline socioeconomic status, including family income and the highest education level among patients, contribute to student clusters of mathematics development. We are also interested in assessing how students' baseline teacher-reported inhibitory control and reading ability account for within-class differences in mathematics development. We standardized the TIC, baseline teacher-reported inhibitory control. For the TVC, reading IRT scores over time, we first calculated the mean and standard deviation of the baseline reading scores and then standardized the score at each wave with the baseline mean value and standard deviation. We built the proposed GMMs with the optimal number of latent classes determined in the enumeration process and provided their estimated likelihood, information criteria, and class-specific residuals in Table \ref{tbl:info}\footnote{In Table \ref{tbl:info}, we also list the estimated likelihood, information criteria, and (class-specific) residual(s) for one-group models with a decomposed TVC and the proposed models with two latent classes. We noticed that the optimal number of latent classes was still three when adding TICs and a decomposed TVC.}.

In addition to the two GMMs with a TVC, we also constructed a reference model with TICs and only the trait effect of the TVC (i.e., a GMM with the standardized baseline socioeconomic variables as the first-type TICs, while the standardized baseline reading scores and teacher-reported inhibitory scores as the second-type TICs). The corresponding model summary is also included in Table \ref{tbl:info}. From the table, we noticed that the inclusion of TICs decreased the estimated likelihood and, therefore, increased AIC and BIC values. Considering the state effects of a TVC reduced the estimated likelihood and raised AIC and BIC values further. However, adding these covariates only slightly affected the class-specific residual variances of mathematics development. Among the two 3-class GMMs with a decomposed TVC, the one with interval-specific changes had the smaller estimated likelihood, AIC, and BIC, indicating that, from the statistical perspective, this model fit the raw data better. 

In Figure \ref{fig:curve_math}, we plot class-specific model-implied curves on the smooth lines of mathematics development for the four 3-class GMMs and examine how adding covariates affects the student clusters and the estimation of the growth factors of the trajectories mathematics IRT scores. We first observed that the estimated mixing proportions of the two proposed GMMs differed from those obtained from the model without the state effects or the model without covariates. Using the GMM with a decomposed TVC into interval-specific slopes as the reference, the Dumenci latent Kappa statistics of the finite mixture model, full mixture model, and the other GMM with a decomposed TVC were $0.26$ with $95\%$ CI ($0.19$, $0.37$), $0.33$ with $95\%$ CI ($0.27$, $0.40$) and $0.98$ with $95\%$ CI ($0.96$, $0.99$), indicating fair, fair and almost perfect agreement. Such findings suggested that adding a TVC in GMMs may change the estimated membership meaningfully, which aligns with the result of the simulation study and that obtained from the application of \citet{Liu2021MoE}. Yet, the decomposition method of a TVC only affected the estimated membership slightly. In addition, we noticed that adding a TVC helped separate the mathematics trajectories of Class 1 and Class 2, shown in Figure \ref{fig:curve_math}. 

The class-specific estimates of the two proposed GMMs are provided in Tables \ref{tbl:TVCslp_est} and \ref{tbl:TVCchg_est}, respectively. The tables show that the mathematics development, TVC, and its trait and state effects are heterogeneous across latent classes. Two GMMs with a decomposed TVC provided similar class-specific estimated trajectories of mathematics development in general, as demonstrated in Tables \ref{tbl:TVCslp_est} and \ref{tbl:TVCchg_est}. In particular, on average, students in Class 3 performed the best in mathematics tests throughout the entire study duration, followed by those in Class 2 and then in Class 1. Mathematics development slowed down around $8$- or $9$- years old for all three classes. In addition, better mathematics performance was associated with better reading ability and better inhibitory control at baseline. As shown in Table \ref{tbl:TVCslp_est}, the estimated mean values of the standardized baseline reading IRT scores/inhibitory control were negative ($-0.67$/$-0.44$), around zero ($-0.26$/$0.10$), and positive ($0.81$/$0.19$) for class 1, Class 2, and Class 3, respectively. Additionally, the slopes of standardized reading scores during the first time interval were $1.51$, $1.96$, and $2.55$ for three latent classes, respectively. These findings suggest that the region with higher reading scores and inhibitory control values was associated with better mathematics development and vice versa. 

\tablehere{5}

\tablehere{6}

The two proposed models' estimated class-specific trait effects of the reading development were also similar. Take the GMM with interval-specific slopes as an example. The trait effects on the initial status of mathematics development were $11.17$, $6.47$, and $6.29$ for the three latent classes, suggesting that the initial mathematics IRT scores would increase $11.17$, $6.47$, and $6.29$ with one SD increase in baseline reading scores, respectively. The estimated class-specific state effects were different between the two models as such effects were based on interval-specific slopes and changes, respectively. In particular, the class-specific state effects were $3.38$, $1.96$, and $1.35$ for the GMM with interval-specific slopes, while they were $8.13$, $3.71$, and $2.76$ for the GMM with interval-specific changes. The interpretation of such state effects is also straightforward. Take the effects of the first class as an example. The effect of $3.38$ suggests that a one-unit increase in the slope of standardized reading IRT scores in a school semester (grade) led to a $3.38$ increase in mathematics final scores that semester (grade) for the students in the first latent class. Similarly, $8.13$ indicates that a one-unit increase in the change of standardized reading scores in a school semester (grade) resulted in a $8.13$ increase in mathematics final scores that semester (grade) for the students in Cluster 1. We also noticed that the estimated growth factors of mathematics development in the first latent class obtained from the GMM with interval-specific changes were relatively smaller than those from the other model. This led to underestimated trajectories indicated by growth factors in the first class, as shown in Figure \ref{fig:curve_TVCchg}. We will discuss this observation in the section Practical Considerations. 

\section*{Discussion}\label{Discussion}
This article further explores the effects on trajectory heterogeneity of a covariate. In particular, it examines two possible ways to decompose a TVC into the trait and state features in the GMM framework. We view the true score of the baseline value of a TVC as the trait feature and evaluate class-specific trait effects on the growth factors of the longitudinal outcome. Moreover, we consider either interval-specific slopes or changes as a set of state features and examine class-specific state effects on the measurements of the longitudinal outcome. We evaluated the proposed GMMs through the simulation study and a real-world application. Generally, the GMMs with a decomposed TVC can separate the longitudinal outcome and estimate the class-specific parameters of interest unbiasedly and accurately with $95\%$ CPs. Similar to other GMMs, increased separation between latent classes improved clustering effects indicated by accuracy values. In addition, we noticed that the mixture models with a TVC, especially when the TVC's state features differed across clusters, improved the accuracy values. It is within our expectations since the inclusion of the TVC increased the separation between latent classes. We also demonstrated the proposed GMMs with real-world data analyses.  

\subsection*{Practical Considerations}\label{D:practical}
This section provides a set of recommendations for empirical researchers based on the simulation study and real-world analysis. \citet{Liu2022decompse} proposed three TVC decomposition methods for the one-group model: (1) a baseline value with interval-specific slopes, (2) a baseline value with interval-specific changes, and (3) a baseline value with changes from baseline at each post-baseline point. All three methods can be extended to the mixture model framework. Therefore, along with the existing GMM-TVC model constructed by \citet{Liu2021MoE}, a TVC can be included in GMMs through four approaches. In the simulation study and the real-world application of the present work, we only considered the first two decomposition methods that are able to provide valid estimates for a one-group model shown by the simulation study in \citet{Liu2022decompse}. It is not our intention to suggest not utilizing the other two approaches since they can also provide insights if the evaluation of the effects of the observed values or change-from-baseline of a TVC is of research interest. 

However, the interpretation of the estimates, especially the parameters related to growth factors of the longitudinal outcome, from the other two methods needs to be cautious. Adding a TVC into growth models allows for regressions of the longitudinal outcome on growth factors and the TVC and, therefore, divides the variability of the observed values at each occasion of the longitudinal outcome into three parts: (1) the variability explained by (class-specific) growth factors that may be further characterized by (second-type) TICs and the trait feature of the TVC, (2) the variability explained by a TVC, and (3) residual variances. As demonstrated by \citet{Liu2022decompse} and the present work, the inclusion of a TVC in growth models usually shrinks the estimates of the (class-specific) growth factors while only slightly affecting the estimates of the residual variances. The greater the state effects and/or the corresponding values of state features are, the higher the squeezing impacts are. Conceptually, the observed values at each point and the changes from baseline at each post-baseline are larger than interval-specific slopes or changes, which leads to more underestimated growth factor means. As we showed in the application section, there are two ways to gauge the gap between the estimated growth factors and those characterized by the raw data. One may construct reference models without a TVC and/or plot the (class-specific) trajectory based on the estimated growth factors and the smooth line in a real-world analysis as shown in Figure \ref{fig:curve_math}. 

Following \citet{Liu2022decompse}, we standardized the TVC with its baseline mean value and standard deviation at each study wave to keep its underlying change patterns and variance-covariance structure and allow for transforming back and obtaining estimates in the original scales. Similar to the findings in \citet{Liu2021MoE}, the clustering algorithm also divides the second-type TIC and the TVC into $K$ regions when separating trajectories of the longitudinal outcome into $K$ latent classes. For example, in our application, standardized baseline inhibitory control and standardized reading IRT scores were split into negative, about zero, and positive regions when classifying mathematics development into three latent classes. This division, along with the heterogeneity in state effects, also suggests that the squeezing impact of the TVC on the estimation of the growth factors may also be different across latent classes. As demonstrated in Figure \ref{fig:curve_TVCchg}, the growth factors in Class 1 might be somewhat underestimated due to relatively greater state effects of the class-specific interval-specific changes and the corresponding feature values. On the contrary, the impacts of the TVC in the other two latent classes were negligible. 

In addition, we still recommend building one-step GMMs with a stepwise fashion for the proposed models, as we demonstrated in the application section. In particular, the enumeration process should be performed to have optimal solutions to the number of latent classes without covariates by following the SEM literature convention. With the stable solution of the number of latent classes, we construct the entire model and estimate parameters of interest. The determination of covariates in mixture models could be data-driven, which can be realized by SEM machine learning techniques, such as EFA and SEM forests, as shown by \citet{Liu2019BLSGMM, Liu2021MoE, Liu2022PBLSGMM}. When applying these machine learning techniques, we recommend only keeping the trait feature of a TVC. 

\subsection*{Methodological Considerations and Future Directions}\label{D:method}
This study has shown several directions in need of further examination. First, a TIC is not allowed to account for between-class differences and within-class heterogeneity simultaneously in this present study following \citet{Liu2021MoE} for both conceptual and technical considerations that have been elaborated in the existing work. However, a TIC, even the trait feature of a TVC, may simultaneously contribute to between- and within-class differences. As demonstrated in the application section, the covariate space of the standardized baseline inhibitory control and reading IRT scores were also divided by the clustering algorithm into the negative, about zero, and positive regions. It suggests that other functions rather than the logistic regression, which only allows for the generalized linear relationship between the covariate and the membership, for the mixing proportions can be developed. One possible function is similar to a tree-based model, and the corresponding estimates based on covariates could be the `boundary' of each region.

Second, as pointed out in the application section, the underestimated growth factors of the longitudinal outcome are due to the squeezing effects of the TVC, which is also a predictor of measurement of the longitudinal outcome at each point. As demonstrated in Figures \ref{fig:curve_TVCslp} and \ref{fig:curve_TVCchg}, such shrinking effects are heterogeneous over time. One may consider relaxing the homogeneous state effects (i.e., $\kappa_{1}$ or $\kappa_{2}$) over time to allow for varying state effects, adjusted by the magnitude of the corresponding state features. Although the investigation of the GMM with a TVC with heterogeneous state effects over time is out of the scope of the current article, it can be a future direction.

Third, as pointed out in earlier sections and summarized in Table \ref{tbl:traj_summary}, the longitudinal outcome may take other functional forms, such as quadratic or negative exponential curves, if the change patterns of the raw data warrant and/or the specific growth coefficients, such as acceleration or growth capacity, are of research interests. The GMMs with a decomposed TVC and the parametric functions listed in Table \ref{tbl:traj_summary} are also provided on the GitHub website for researchers interested in employing them. The TVC can also take other functional forms with the specification proposed by \citet{Liu2022LCSM}, which allows for the examination of growth coefficients of the TVC in addition to those interval-specific slopes or changes. 

Fourth, the proposed GMMs are illustrated with the same time structure for the longitudinal outcome and the TVC. Yet, one may extend the models to analyze a process with longitudinal variables having different measurement times since the models are constructed in the framework of individual measurement occasions. Last, the proposed models can also be utilized to analyze data with dropout under the assumptions of missing at random since the FIML technique is employed for model estimation. 

\subsection*{Concluding Remarks}\label{D:conclude}
This article proposes two GMMs with a decomposed TVC to evaluate the trait effect and state effect of the TVC on sample heterogeneity separately. In the two models, we view the baseline values of the TVC as its trait feature and the interval-specific slopes or changes as a collection of state features. The proposed growth models allow for the examination of the heterogeneity in the trait and state features and their heterogeneous effects. They can be further extended in practice and further investigated in methodology. 

\bibliographystyle{apalike}
\bibliography{Extension10}

\renewcommand{\thesection}{Appendix \Alph{section}}
\renewcommand{\thesubsection}{A.\arabic{subsection}}
\setcounter{section}{0}
\section{Formula Derivation}
\subsection{Class-specific Model-implied Mean Vector and Variance-covariance Matrix of the Longitudinal Variables}\label{supp1}
\renewcommand{\theequation}{A.\arabic{equation}}
\setcounter{equation}{0}
For the GMMs with TICs and a TVC that is decomposed into the baseline value and interval-specific slopes, the class-specific model-implied mean vector and variance-covariance matrix of the TVC and the longitudinal outcome ($\boldsymbol{y}_{i}$) for the $i^{th}$ individual can be expressed as
\begin{equation}\nonumber
\boldsymbol{\mu}^{(k)}_{i}=\begin{pmatrix}
\boldsymbol{\mu}^{(k)[x]}_{i} \\ \boldsymbol{\mu}^{(k)[y]}_{i} 
\end{pmatrix}=\begin{pmatrix}
\boldsymbol{\Lambda}^{(k)[x]}_{i} & \boldsymbol{0} \\ \boldsymbol{0} & \boldsymbol{\Lambda}^{(k)[y]}_{i}
\end{pmatrix}\times\begin{pmatrix}
\boldsymbol{\mu}^{(k)[x]}_{\boldsymbol{\eta}} \\ \boldsymbol{\mu}^{(k)[y]}_{\boldsymbol{\eta}}
\end{pmatrix}+\kappa^{(k)}_{1}\times\begin{pmatrix} \boldsymbol{0} \\ \boldsymbol{\mu}^{(k)}_{dx} \end{pmatrix},
\end{equation}
and
\begin{align}
\boldsymbol{\Sigma}^{(k)}_{i}&=\begin{pmatrix}
\boldsymbol{\Sigma}^{(k)[x]}_{i} & \boldsymbol{\Sigma}^{(k)[xy]}_{i} \\ & \boldsymbol{\Sigma}^{(k)[y]}_{i}
\end{pmatrix} \nonumber\\
&=\begin{pmatrix}
\boldsymbol{\Lambda}^{(k)[x]}_{i} & \boldsymbol{0} \\ \boldsymbol{0} & \boldsymbol{\Lambda}^{(k)[y]}_{i}
\end{pmatrix}\times \begin{pmatrix}
\boldsymbol{\Phi}^{[x]}_{\boldsymbol{\eta}} & \boldsymbol{0} \\
\boldsymbol{0} & \boldsymbol{\text{Var(y)}^{(k)}}
\end{pmatrix} \times\begin{pmatrix}
\boldsymbol{\Lambda}^{(k)[x]}_{i} & \boldsymbol{0} \\ \boldsymbol{0} & \boldsymbol{\Lambda}^{(k)[y]}_{i}
\end{pmatrix}^{T}\nonumber\\
&+\begin{pmatrix}\boldsymbol{0} & \boldsymbol{0} \\ \boldsymbol{0} & \kappa^{(k)2}_{1}\boldsymbol{\phi}^{(k)}_{dx}
\end{pmatrix}+\begin{pmatrix} \theta^{(k)[x]}_{\epsilon}\boldsymbol{I} &
\theta^{(k)[xy]}_{\epsilon}\boldsymbol{I} \\ & \theta^{(k)[y]}_{\epsilon}\boldsymbol{I} \end{pmatrix}, \nonumber
\end{align}
respectively, where $\boldsymbol{\mu}^{(k)}_{dx}$ and $\boldsymbol{\phi}^{(k)}_{dx}$ are the class-specific means and variances of the interval-specific slopes, which are not freely estimated parameters in the proposed GMMs. With the function \textit{mxAlgebra()} in \textit{OpenMx}, these non-estimable parameters are often created, and the point estimates are stored in the corresponding OpenMx objective. Their standard errors are usually obtained by the function \textit{mxSE()}. In addition, $\boldsymbol{\mu}^{(k)[y]}_{\boldsymbol{\eta}}$ and $\boldsymbol{\text{Var(y)}}^{(k)}$ are the class-specific conditional mean vector and variance-covariance matrix of the growth factors of the longitudinal outcome on the second-type TIC and the true score of the TVC initial status, which can be further expressed as
\begin{equation}\nonumber
\boldsymbol{\mu}^{(k)[y]}_{\boldsymbol{\eta}}=\boldsymbol{\alpha}^{(k)[y]}+\begin{pmatrix}\boldsymbol{\beta^{(k)}_{\text{TIC}}} & \boldsymbol{\beta^{(k)}_{\text{TVC}}}\end{pmatrix}\times\begin{pmatrix}\mu^{(k)}_{x} \\ \mu^{(k)[x]}_{\eta_{0}}\end{pmatrix},
\end{equation}
and
\begin{equation}\nonumber
\boldsymbol{\text{Var(y)}}^{(k)}=\boldsymbol{\Psi}^{(k)[y]}_{\boldsymbol{\eta}}+\begin{pmatrix}\boldsymbol{\beta^{(k)}_{\text{TIC}}} & \boldsymbol{\beta^{(k)}_{\text{TVC}}}\end{pmatrix}\times\begin{pmatrix} \phi^{(k)}_{x} & \rho^{(k)}_{\text{BL}}\sqrt{\phi^{(k)}_{x}\phi^{(k)[x]}_{00}} \\
\rho^{(k)}_{\text{BL}}\sqrt{\phi^{(k)}_{x}\phi^{(k)[x]}_{00}} & \phi^{(k)[x]}_{00}
\end{pmatrix}\times\begin{pmatrix}\boldsymbol{\beta^{(k)}_{\text{TIC}}} & \boldsymbol{\beta^{(k)}_{\text{TVC}}}\end{pmatrix}^{T},
\end{equation}
respectively, where $\mu^{(k)[x]}_{\eta_{0}}$ and $\phi^{(k)[x]}_{00}$ are the class-specific mean and variance of the TVC initial status, and $\rho^{(k)}_{\text{BL}}$ is the class-specific correlation between the second-type TIC and the true value of the TVC initial status.

For the GMMs with TICs and a TVC that is decomposed into the baseline value and interval-specific changes, the class-specific model-implied mean vector and variance-covariance matrix of the TVC and the longitudinal outcome ($\boldsymbol{y}_{i}$) for the $i^{th}$ individual can be expressed as
\begin{equation}\nonumber
\boldsymbol{\mu}^{(k)}_{i}=\begin{pmatrix}
\boldsymbol{\mu}^{(k)[x]}_{i} \\ \boldsymbol{\mu}^{(k)[y]}_{i} 
\end{pmatrix}=\begin{pmatrix}
\boldsymbol{\Lambda}^{(k)[x]}_{i} & \boldsymbol{0} \\ \boldsymbol{0} & \boldsymbol{\Lambda}^{(k)[y]}_{i}
\end{pmatrix}\times\begin{pmatrix}
\boldsymbol{\mu}^{(k)[x]}_{\boldsymbol{\eta}} \\ \boldsymbol{\mu}^{(k)[y]}_{\boldsymbol{\eta}}
\end{pmatrix}+\kappa^{(k)}_{2}\times\begin{pmatrix} \boldsymbol{0} \\ \boldsymbol{\mu}^{(k)}_{\delta x} \end{pmatrix},
\end{equation}
and
\begin{align}
\boldsymbol{\Sigma}^{(k)}_{i}&=\begin{pmatrix}
\boldsymbol{\Sigma}^{(k)[x]}_{i} & \boldsymbol{\Sigma}^{(k)[xy]}_{i} \\ & \boldsymbol{\Sigma}^{(k)[y]}_{i}
\end{pmatrix} \nonumber\\
&=\begin{pmatrix}
\boldsymbol{\Lambda}^{(k)[x]}_{i} & \boldsymbol{0} \\ \boldsymbol{0} & \boldsymbol{\Lambda}^{(k)[y]}_{i}
\end{pmatrix}\times \begin{pmatrix}
\boldsymbol{\Phi}^{[x]}_{\boldsymbol{\eta}} & \boldsymbol{0} \\
\boldsymbol{0} & \boldsymbol{\text{Var(y)}^{(k)}}
\end{pmatrix} \times\begin{pmatrix}
\boldsymbol{\Lambda}^{(k)[x]}_{i} & \boldsymbol{0} \\ \boldsymbol{0} & \boldsymbol{\Lambda}^{(k)[y]}_{i}
\end{pmatrix}^{T}\nonumber\\
&+\begin{pmatrix}\boldsymbol{0} & \boldsymbol{0} \\ \boldsymbol{0} & \kappa^{(k)2}_{2}\boldsymbol{\phi}^{(k)}_{\delta x}
\end{pmatrix}+\begin{pmatrix} \theta^{(k)[x]}_{\epsilon}\boldsymbol{I} &
\theta^{(k)[xy]}_{\epsilon}\boldsymbol{I} \\ & \theta^{(k)[y]}_{\epsilon}\boldsymbol{I} \end{pmatrix}, \nonumber
\end{align}
respectively, where $\boldsymbol{\mu}^{(k)}_{\delta x}$ and $\boldsymbol{\phi}^{(k)}_{\delta x}$ are the class-specific means and variances of the interval-specific changes, which are not freely estimable parameters and can be derived using the function \textit{mxAlgebra()}. $\boldsymbol{\mu}^{(k)[y]}_{\boldsymbol{\eta}}$ and $\boldsymbol{\text{Var(y)}}^{(k)}$ have the same definitions as the previous equations. 

\newpage


\renewcommand\thefigure{\arabic{figure}}
\setcounter{figure}{0}

\begin{figure}
\centering
\begin{subfigure}{.5\textwidth}
  \centering
  \includegraphics[width=1\linewidth]{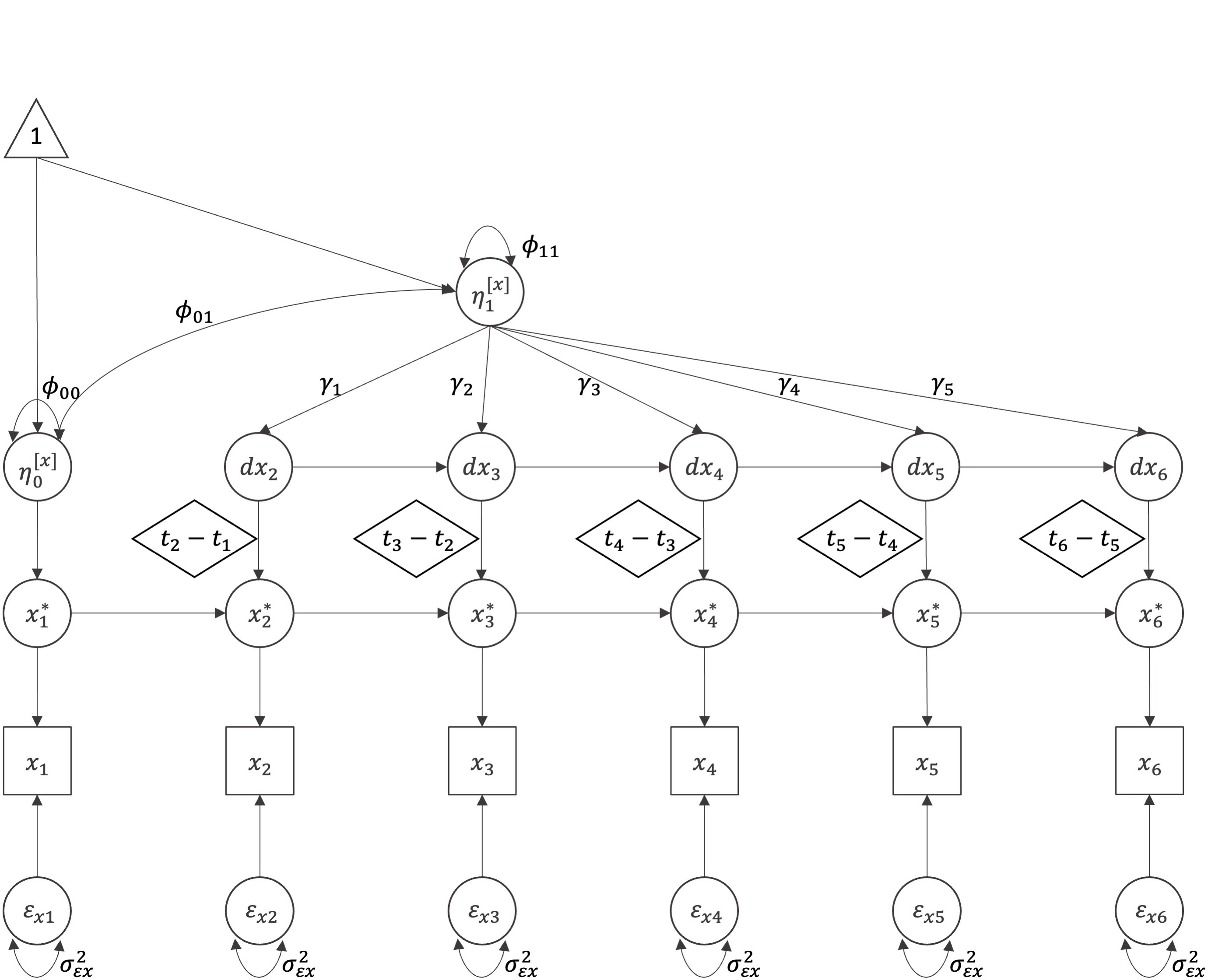}
  \caption{Estimating Interval-specific Slopes}
  \label{fig:path_slp}
\end{subfigure}%
\begin{subfigure}{.5\textwidth}
  \centering
  \includegraphics[width=1\linewidth]{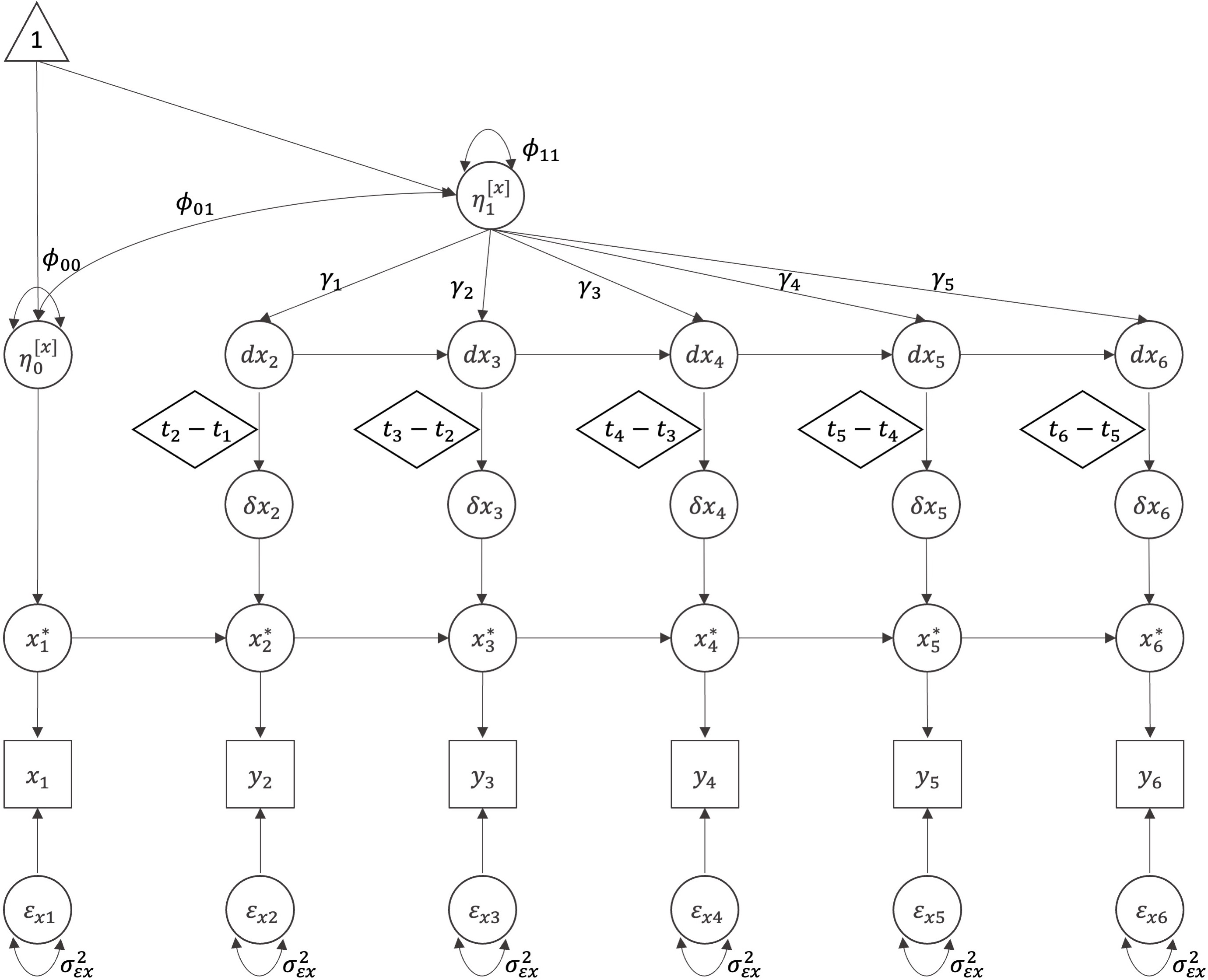}
  \caption{Estimating Interval-specific Changes}
  \label{fig:path_chg}
\end{subfigure}
\caption{Two Possible Path Diagram to Decompose Time-varying Covariates\\
Note: boxes=manifested variables, circles=latent variables, single arrow=regression paths;
doubled arrow=(co)variances; triangle=constant; diamonds=definition variables.\\
In both models, $\gamma_{1}$ is set as $1$ for model identification considerations.}
\label{fig:path_ext}
\end{figure}

\begin{figure}
\centering
\begin{subfigure}{.5\textwidth}
  \centering
  \includegraphics[width=1\linewidth]{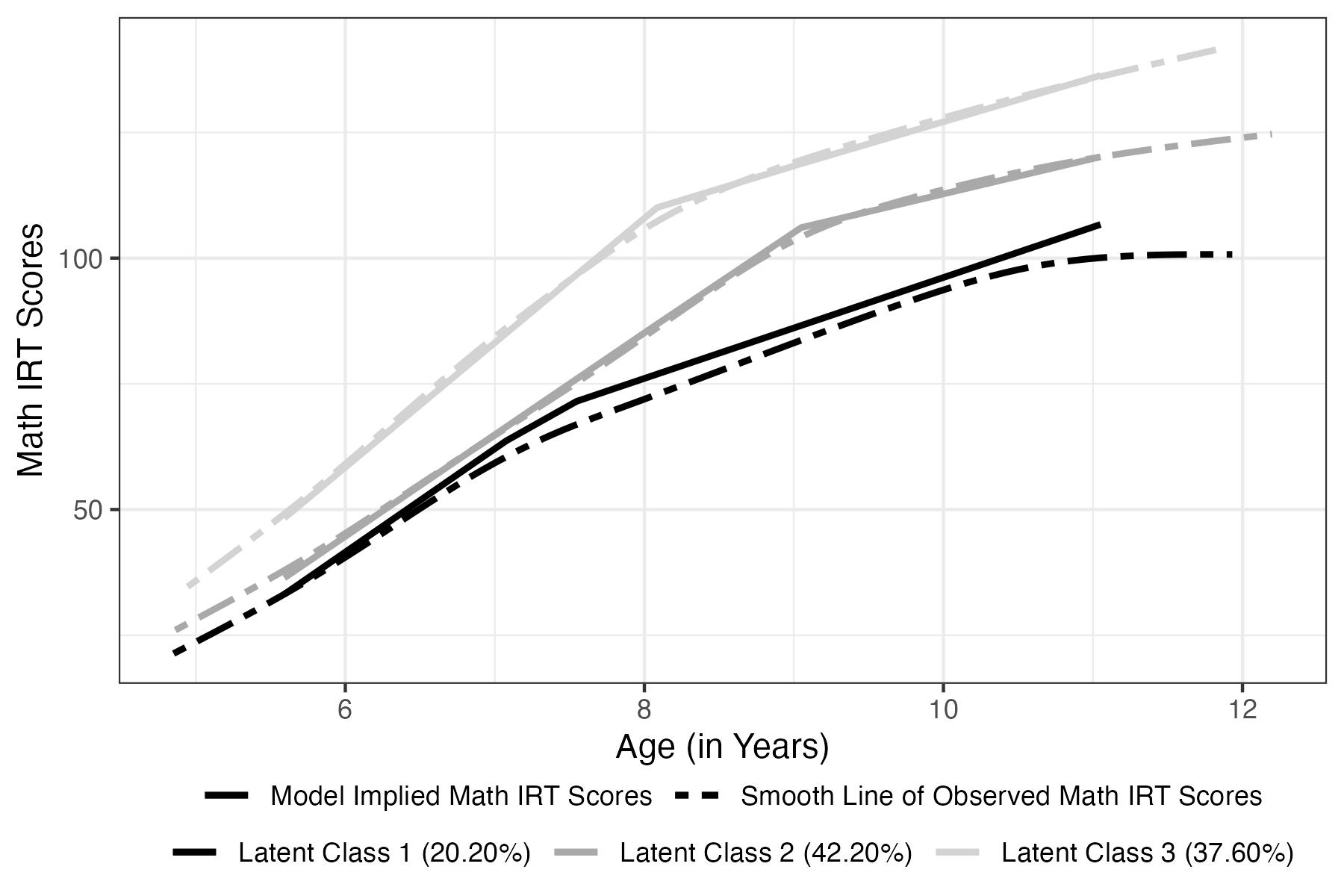}
  \caption{Model 1}
  \label{fig:curve_FMM}
\end{subfigure}%
\begin{subfigure}{.5\textwidth}
  \centering
  \includegraphics[width=1\linewidth]{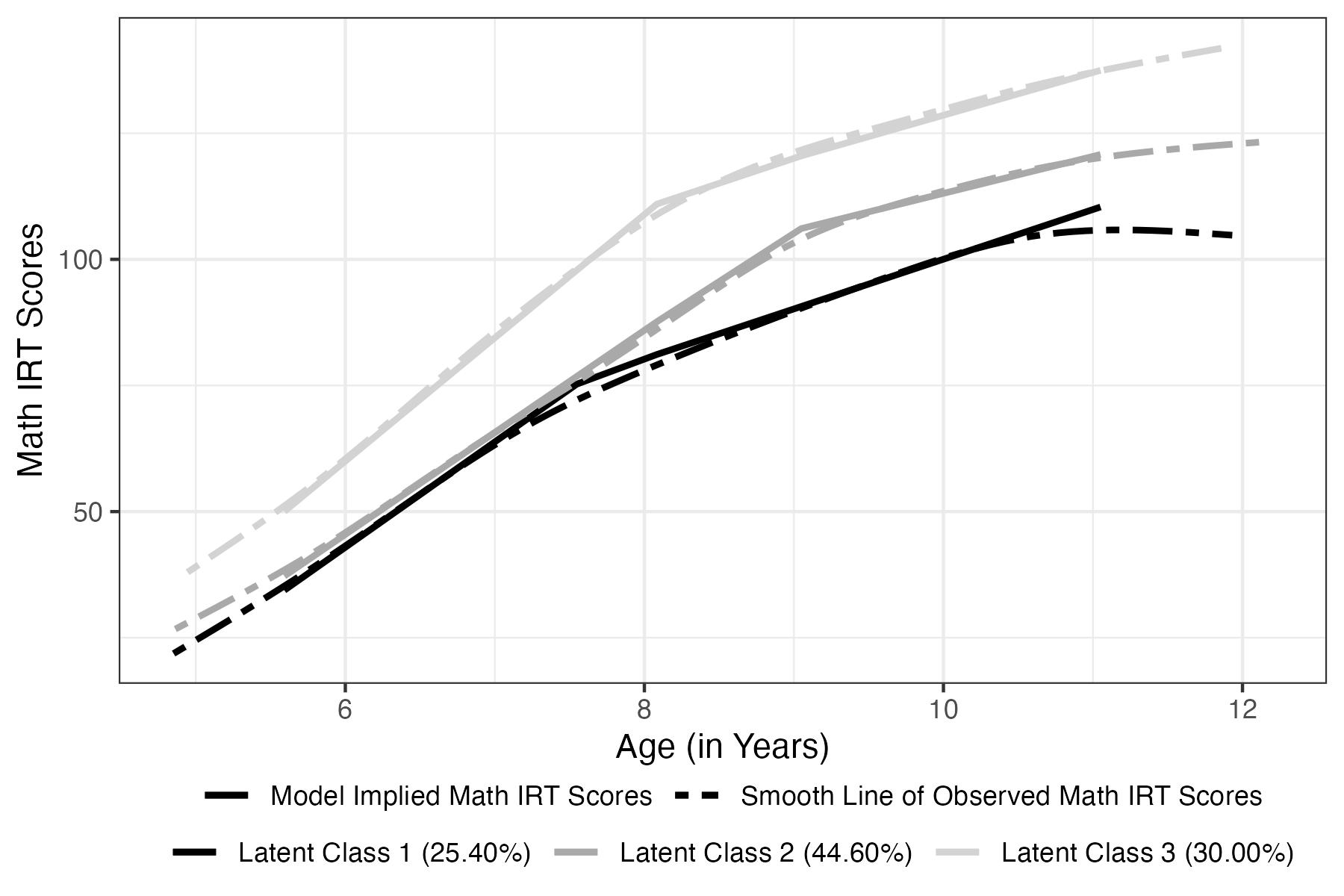}
  \caption{Model 2}
  \label{fig:curve_Full}
\end{subfigure}
\vskip\baselineskip
\begin{subfigure}{.5\textwidth}
  \centering
  \includegraphics[width=1\linewidth]{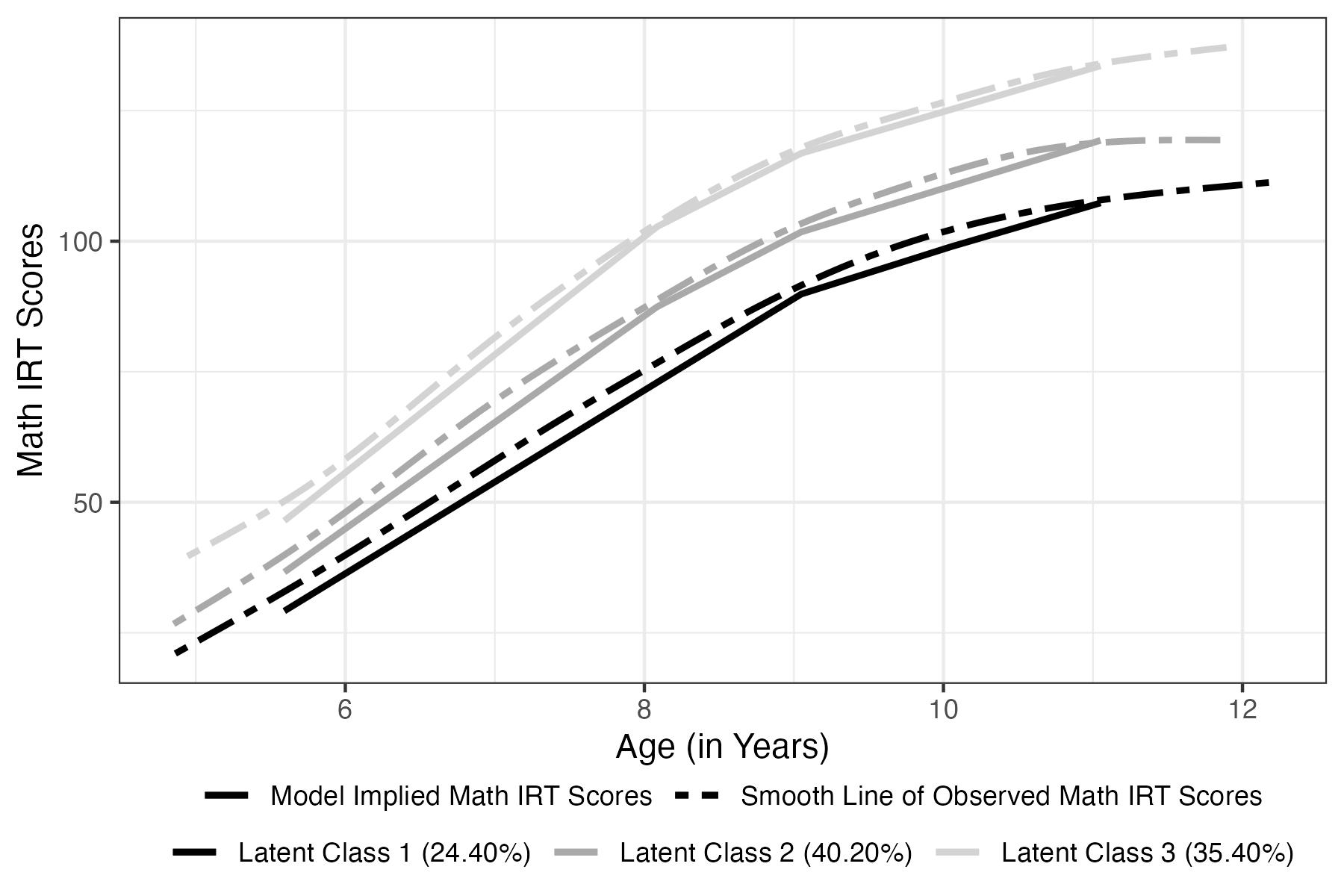}
  \caption{Model 3}
  \label{fig:curve_TVCslp}
\end{subfigure}%
\begin{subfigure}{.5\textwidth}
  \centering
  \includegraphics[width=1\linewidth]{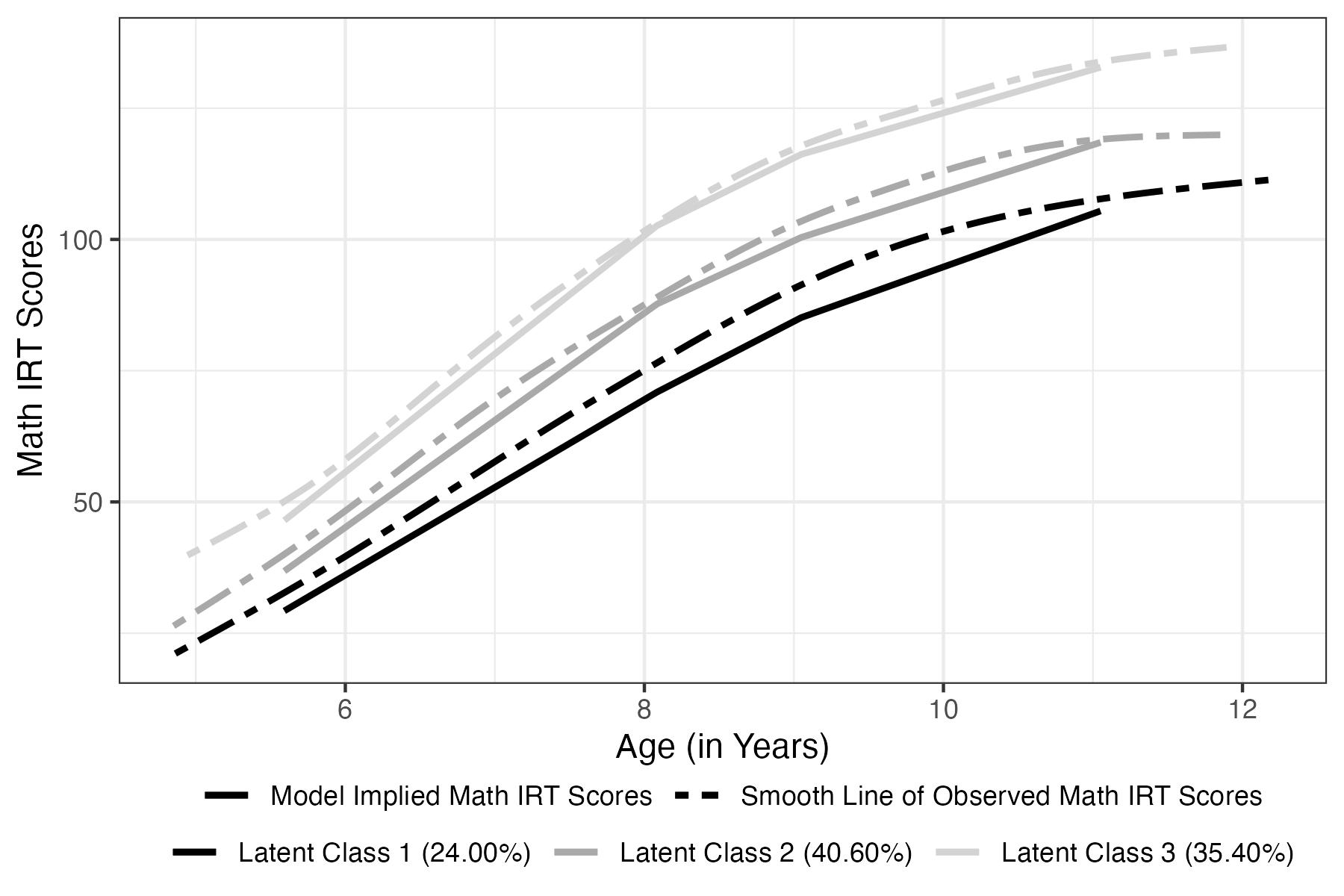}
  \caption{Model 4}
  \label{fig:curve_TVCchg}
\end{subfigure}
\caption{Class-specific Growth-Factors Implied Trajectory and Smooth Line of Mathematics Performance\\
Note: Model 1 is the growth mixture model without covariates; Model 2 is the growth mixture model with two first-type covariates, one second-type covariate, and TVC baseline; Model 3 is the growth mixture model with two first-type covariates, one second-type covariate, and a decomposed TVC with interval-specific slopes; Model 4 is the growth mixture model with two first-type covariates, one second-type covariate, and a decomposed TVC with interval-specific changes.}
\label{fig:curve_math}
\end{figure}


\renewcommand\thetable{\arabic{table}}
\setcounter{table}{0}

\begin{table}
\centering
\footnotesize
\rotatebox{90}{%
\begin{threeparttable}
\setlength{\tabcolsep}{5pt}
\renewcommand{\arraystretch}{0.75}
\caption{Overview of Linear and Four Common Nonlinear Functional Forms in Growth Mixture Models}
\begin{tabular}{p{6cm}p{5cm}p{12cm}}
\hline
\hline
\textbf{Growth Factors}\tnote{a} & \textbf{Factor Loadings}\tnote{b} & \textbf{Interpretation of Growth Coefficients} \\
\hline
\multicolumn{3}{l}{\textbf{Linear Function}: $y_{ij}=\eta^{[y]}_{0i}+\eta^{[y]}_{1i}\times{t_{ij}}+\epsilon^{[y]}_{ij}$}\\
\hline
\multirow{2}{*}{$\boldsymbol{\eta}^{[y]}_{i}=\begin{pmatrix}\eta^{[y]}_{0i} & \eta^{[y]}_{1i}\end{pmatrix}$} & 
\multirow{2}{*}{$\boldsymbol{\Lambda}^{[y]}_{i}=\begin{pmatrix}1 & t_{ij} \end{pmatrix}$} & 
$\eta^{[y]}_{0i}$: the initial status (i.e., baseline value) \\
& & $\eta^{[y]}_{1i}$: the slope \\
\hline
\hline
\multicolumn{3}{l}{\textbf{Quadratic Function}: $y_{ij}=\eta^{[y]}_{0i}+\eta^{[y]}_{1i}\times{t_{ij}}+\eta^{[y]}_{2i}\times{t^{2}_{ij}}+\epsilon^{[y]}_{ij}$}\\
\hline
\multirow{3}{*}{$\boldsymbol{\eta}^{[y]}_{i}=\begin{pmatrix}\eta^{[y]}_{0i} & \eta^{[y]}_{1i} & \eta^{[y]}_{2i}\end{pmatrix}$} & 
\multirow{3}{*}{$\boldsymbol{\Lambda}^{[y]}_{i}=\begin{pmatrix}1 & t_{ij} & t^{2}_{ij} \end{pmatrix}$} & 
$\eta^{[y]}_{0i}$: the initial status (i.e., baseline value) \\
& & $\eta^{[y]}_{1i}$: the linear slope \\
& & $\eta^{[y]}_{2i}$: the quadratic slope \\
\hline
\hline
\multicolumn{3}{l}{\textbf{Negative Exponential Function}: $y_{ij}=\eta^{[y]}_{0i}+\eta^{[y]}_{1i}\times(1-\exp(-b\times {t_{ij}}))+\epsilon^{[y]}_{ij}$}\\
\hline
\multirow{3}{*}{$\boldsymbol{\eta}^{[y]}_{i}=\begin{pmatrix}\eta^{[y]}_{0i} & \eta^{[y]}_{1i} \end{pmatrix}$} & 
\multirow{3}{*}{$\boldsymbol{\Lambda}^{[y]}_{i}=\begin{pmatrix}1 & 1-\exp(-b\times {t_{ij}}) \end{pmatrix}$} & 
$\eta^{[y]}_{0i}$: the initial status (i.e., baseline value) \\
& & $\eta^{[y]}_{1i}$: the change from baseline value to asymptotic level \\
& & $b$\tnote{c}: the log-ratio of rate-of-change at $t_{ij}$ to that at $t_{i(j-1)}$ \\
\hline
\hline
\multicolumn{3}{l}{\textbf{Jenss-Bayley Function}: $y_{ij}=\eta^{[y]}_{0i}+\eta^{[y]}_{1i}\times t_{ij}+\eta^{[y]}_{2i}\times(\exp(c\times t_{ij})-1)+\epsilon^{[y]}_{ij}$}\\
\hline
\multirow{4}{*}{$\boldsymbol{\eta}^{[y]}_{i}=\begin{pmatrix}\eta^{[y]}_{0i} & \eta^{[y]}_{1i} & \eta^{[y]}_{2i} \end{pmatrix}$} & 
\multirow{4}{*}{$\boldsymbol{\Lambda}^{[y]}_{i}=\begin{pmatrix}1 & t_{ij} & \exp(c\times {t_{ij}}-1) \end{pmatrix}$} & 
$\eta^{[y]}_{0i}$: the initial status (i.e., baseline value) \\
& & $\eta^{[y]}_{1i}$: the slope of linear asymptote \\
& & $\eta^{[y]}_{2i}$: the change from baseline value to the linear asymptote intercept\\
& & $c$\tnote{c}: the log-ratio of acceleration at $t_{ij}$ to that at $t_{i(j-1)}$ \\
\hline
\hline
\multicolumn{3}{l}{\textbf{The Linear-linear Function with a Fixed Knot}\tnote{d}: $y_{ij}=\begin{cases}
\eta^{[y]}_{0i}+\eta^{[y]}_{1i}\times t_{ij}+\epsilon^{[y]}_{ij}, &  t_{ij}<\gamma \\
\eta^{[y]}_{0i}+\eta^{[y]}_{1i}\times \gamma+\eta^{[y]}_{2i}\times(t_{ij}-\gamma)+\epsilon^{[y]}_{ij}, & t_{ij}\ge\gamma \\
\end{cases}$}\\
\hline
\multirow{4}{*}{$\boldsymbol{\eta}^{[y]'}_{i}=\begin{pmatrix}\eta^{[y]}_{0i}+\gamma\eta^{[y]}_{1i} & \frac{\eta^{[y]}_{1i}+\eta^{[y]}_{2i}}{2} & \frac{\eta^{[y]}_{2i}-\eta^{[y]}_{1i}}{2}\end{pmatrix}$} & 
\multirow{4}{*}{$\boldsymbol{\Lambda}^{[y]'}_{i}=\begin{pmatrix}1 & t_{ij}-\gamma & |t_{ij}-\gamma| \end{pmatrix}$} & 
$\eta^{[y]}_{0i}$: the initial status (i.e., baseline value) \\
& & $\eta^{[y]}_{1i}$: the slope of the first linear segment \\
& & $\eta^{[y]}_{2i}$: the slope of the second linear segment \\
& & $\gamma$\tnote{c}: the knot (i.e., transition time from $1^{st}$ linear segment to $2^{nd}$ linear segment) \\
\hline
\hline
\end{tabular}
\label{tbl:traj_summary}
\begin{tablenotes}
\small
\item[a] {Growth factors ($\boldsymbol{\eta}^{[y]}_{i}$) is a $C\times 1$ vector, where $C$ is the number of growth factors.}\\
\item[b] {Factor loadings ($\boldsymbol{\Lambda}^{[y]}_{i}$) is a $J\times C$ matrix, where $J$ is the number of repeated measurements while $C$ is the number of growth factors.}\\
\item[c] {The coefficient $b$, $c$, and $\gamma$ are assumed to be the same across the population in the table. They can be viewed as additional growth factors in the corresponding functional forms, for which we can estimate their mean values and variances. The construction of growth models with individual levels $b$, $c$, or $\gamma$ in the structural equation modeling framework requires `linearization', which can be realized through the Taylor series expansion. Multiple earlier studies, such as \citet{Preacher2015repara}, \citet[Chapter~11]{Grimm2016growth}, and \citet{Liu2019BLSGM}, have well-documented technical details for `linearization'.}\\
\item[d] {The linear-linear functional form (also referred to as the bilinear function) is widely used to depict longitudinal processes in multiple areas, including short- and long-term recovery periods in psychotherapy or earlier- and later-stages of intellectual development. The knot from the first to the second stage (i.e., $\gamma$) can be estimated as a free parameter. The reparameterized growth factors (i.e., $\boldsymbol{\eta}^{[y]'}_{i}$) and the corresponding factor loadings (i.e., $\boldsymbol{\Lambda}^{[y]'}_{i}$) shown in the table enable one to unify the piecewise function and fit the model in the structural equation modeling framework. Technical details are documented in \citet{Liu2019BLSGM}.} 
\end{tablenotes}
\end{threeparttable}}%
\end{table}

\begin{table}
\centering
\begin{threeparttable}
\caption{Definitions and Estimators of Metrics to Evaluate an Estimate $\hat{\theta}$ of Parameter $\theta$}
\begin{tabular}{p{4cm}p{4.5cm}p{6cm}}
\hline
\hline
\textbf{Criteria} & \textbf{Definition} & \textbf{Estimator} \\
\hline
Relative Bias & $E_{\hat{\theta}}(\hat{\theta}-\theta)/\theta$ & $\sum_{s=1}^{S}(\hat{\theta}_{s}\tnote{a}-\theta)/\theta S\tnote{b}$ \\
Empirical SE & $\sqrt{Var(\hat{\theta})}$ & $\sqrt{\sum_{s=1}^{S}(\hat{\theta}_{s}-\bar{\theta}\tnote{c})^{2}/(S-1)}$ \\
Relative RMSE & $\sqrt{E_{\hat{\theta}}(\hat{\theta}-\theta)^{2}}/\theta$ & $\sqrt{\sum_{s=1}^{S}(\hat{\theta}_{s}-\theta)^{2}/S}/\theta$ \\
Coverage Probability & $Pr(\hat{\theta}_{\text{lower}}\le\theta\le\hat{\theta}_{\text{upper}})$ & $\sum_{s=1}^{S}I(\hat{\theta}_{\text{lower},s}\le\theta\le\hat{\theta}_{\text{upper},s})\tnote{d}/S$\\
\hline
\hline
\end{tabular}
\label{tbl:metric}
\begin{tablenotes}
\small
\item[a] {$\hat{\theta}_{s}$: the estimate of $\theta$ from the $s^{th}$ replication.}\\ 
\item[b] {$S$: the number of replications, which is set as $1,000$ in the simulation study.}\\
\item[c] {$\bar{\theta}$: the mean of $\hat{\theta}_{s}$'s across all $1,000$ replications.}\\
\item[d] {$I()$ is an indicator function.}
\end{tablenotes}
\end{threeparttable}
\end{table}

\begin{table}
\centering
\resizebox{1.15\textwidth}{!}{
\begin{threeparttable}
\setlength{\tabcolsep}{4pt}
\renewcommand{\arraystretch}{0.6}
\caption{Simulation Design for GMMs with TICs and a Decomposed TVC with Individual Measurement Occasions}
\begin{tabular}{p{4.6cm} p{13.6cm}}
\hline
\hline
\multicolumn{2}{c}{\textbf{Sample Size and common conditions across longitudinal processes}}\\
\hline
Sample size & $n=500$ \\
\hline
Study wave ($t_{j}$) & $10$ equally-spaced: $t_{j}=0, 1.00, \dots, J-1\quad(J=10)$\\
\hline
Individual time ($t_{ij}$) & $t_{ij} \sim U(t_{j}-\Delta, t_{j}+\Delta)$ ($\Delta=0.25$) \\
\hline
\multirow{2}{*}{Logistic coefficients} & $\beta_{0}=0$ (allocation ratio is about 1:1), $\beta_{1}=\log(1.5)$, $\beta_{2}=\log(1.7)$\\
& $\beta_{0}=0.775$ (allocation ratio is about 1:2), $\beta_{1}=\log(1.5)$, $\beta_{2}=\log(1.7)$\\
\hline
Mahalanobis distance & $d=0.86$ \\
\hline
\multirow{2}{*}{First-type covariates} & $\boldsymbol{x}_{gi}\sim N(\boldsymbol{0}, \begin{pmatrix} 1.0 & 0.3 \\ 0.3 & 1.0 \end{pmatrix})$ \\
\hline
\hline
\multicolumn{2}{c}{\textbf{Class-specific parameters of the growth curves of the longitudinal outcome}}\\
\hline
\hline
\multirow{2}{*}{Mean vector} & $\begin{pmatrix} \mu^{(1)[y]}_{\eta_{0}} & \mu^{(1)[y]}_{\eta_{1}} & \mu^{(1)[y]}_{\eta_{2}} \end{pmatrix}=\begin{pmatrix} 48 & 4.5 & 1.65 \end{pmatrix}$\\
& $\begin{pmatrix} \mu^{(2)[y]}_{\eta_{0}} & \mu^{(2)[y]}_{\eta_{1}} & \mu^{(2)[y]}_{\eta_{2}} \end{pmatrix}=\begin{pmatrix} 52 & 5.0 & 1.80 \end{pmatrix}$\\
\hline
\multirow{3}{*}{Variance-covariance matrix} & $\begin{pmatrix} \psi^{(1)[y]}_{00} & \psi^{(1)[y]}_{01} & \psi^{(1)[y]}_{02} \\ \psi^{(1)[y]}_{01} & \psi^{(1)[y]}_{11} & \psi^{(1)[y]}_{12} \\ \psi^{(1)[y]}_{02} & \psi^{(1)[y]}_{12} & \psi^{(1)[y]}_{22} \end{pmatrix}=\begin{pmatrix} \psi^{(2)[y]}_{00} & \psi^{(2)[y]}_{01} & \psi^{(2)[y]}_{02} \\ \psi^{(2)[y]}_{01} & \psi^{(2)[y]}_{11} & \psi^{(2)[y]}_{12} \\ \psi^{(2)[y]}_{02} & \psi^{(2)[y]}_{12} & \psi^{(2)[y]}_{22} \end{pmatrix}=\begin{pmatrix} 25 & 1.5 & 1.5 \\
1.5 & 1.0 & 0.3 \\ 1.5 & 0.3 & 1.0 \end{pmatrix}$\\
\hline
\multirow{3}{*}{Knot location} & $\gamma^{(1)}=5.00$, $\gamma^{(2)}=4.00$ \\
& $\gamma^{(1)}=5.25$, $\gamma^{(2)}=3.75$ \\ & $\gamma^{(1)}=5.50$, $\gamma^{(2)}=3.50$ \\
\hline
Residual variance & $\theta^{(1)[y]}_{\epsilon}=\theta^{(2)[y]}_{\epsilon}=1$ or $2$ \\
\hline
\hline
\multicolumn{2}{c}{\textbf{Class-specific parameters of the growth curves of the time-varying covariate}}\\
\hline
\hline
\textbf{Variables} & \textbf{Conditions} \\
\hline
\multirow{3}{*}{Mean vector} & Scenario 1: $\begin{pmatrix} \mu^{(1)[x]}_{\eta_{0}} & \mu^{(1)[x]}_{\eta_{1}} \end{pmatrix}=\begin{pmatrix} 0 & 5 \end{pmatrix}$, $\begin{pmatrix} \mu^{(2)[x]}_{\eta_{0}} & \mu^{(2)[x]}_{\eta_{1}} \end{pmatrix}=\begin{pmatrix} 0 & 5 \end{pmatrix}$\\
& Scenario 2: $\begin{pmatrix} \mu^{(1)[x]}_{\eta_{0}} & \mu^{(1)[x]}_{\eta_{1}} \end{pmatrix}=\begin{pmatrix} 0 & 4 \end{pmatrix}$, $\begin{pmatrix} \mu^{(2)[x]}_{\eta_{0}} & \mu^{(2)[x]}_{\eta_{1}} \end{pmatrix}=\begin{pmatrix} 0 & 6 \end{pmatrix}$\\
& Scenario 3: $\begin{pmatrix} \mu^{(1)[x]}_{\eta_{0}} & \mu^{(1)[x]}_{\eta_{1}} \end{pmatrix}=\begin{pmatrix} 0 & 5 \end{pmatrix}$, $\begin{pmatrix} \mu^{(2)[x]}_{\eta_{0}} & \mu^{(2)[x]}_{\eta_{1}} \end{pmatrix}=\begin{pmatrix} 0 & 5 \end{pmatrix}$\\
\hline
\multirow{2}{*}{Variance-covariance matrix} & $\begin{pmatrix} \psi^{(1)[x]}_{00} & \psi^{(1)[x]}_{01} \\ \psi^{(1)[x]}_{01} & \psi^{(1)[x]}_{11} \end{pmatrix}=\begin{pmatrix} \psi^{(2)[x]}_{00} & \psi^{(2)[x]}_{01} \\ \psi^{(2)[x]}_{01} & \psi^{(2)[x]}_{11} \end{pmatrix}=\begin{pmatrix} 1.0 & 0.3 \\
0.3 & 1.0 \end{pmatrix}$\\
\hline
\multirow{4}{*}{Relative Rate-of-Change\tnote{a}} & Scenario 1: $\gamma^{(1)}=\gamma^{(2)}=1.0/0.9/0.8/0.7/0.6/0.5/0.4/0.3/0.2$ \\
& Scenario 2: $\gamma^{(1)}=\gamma^{(2)}=1.0/0.9/0.8/0.7/0.6/0.5/0.4/0.3/0.2$ \\
& Scenario 3: $\gamma^{(1)}=1.0/0.88/0.76/0.64/0.52/0.40/0.28/0.16/0.04$ and \\ & $\gamma^{(2)}=1.0/0.92/0.84/0.76/0.68/0.60/0.52/0.44/0.36$ \\
\hline
Residual variance & $\theta^{(1)[x]}_{\epsilon}=\theta^{(2)[x]}_{\epsilon}=1$ \\
\hline
\hline
\multicolumn{2}{c}{\textbf{Class-specific parameters related to second-type time-invariant covariate and trait feature}}\\
\hline
\hline
\textbf{Variables} & \textbf{Conditions} \\
\hline
Correlation\tnote{b} & $\rho^{(1)}_{BL}=\rho^{(2)}_{BL}=0.3$ \\
\hline
\multirow{4}{*}{Coef. to growth factors} & Class 1: Trait feature explains $14\%$ variability of growth factors\tnote{b}, and \\ & Time-invariant covariate explains $6\%$ variability of growth factors\\
\cline{2-2}
& Class 2: Trait feature explains $7\%$ variability of growth factors\tnote{b}, and \\& Time-invariant covariate explains $3\%$ variability of growth factors \\ 
\hline
Time-invariant covariate & $x_{ei}|(z_{i}=1)\sim N(0, 1^{2})$; $x_{ei}|(z_{i}=2)\sim N(0, 1^{2})$ \\
\hline
\hline
\multicolumn{2}{c}{\textbf{Parameters related state feature}}\\
\hline
\hline
\textbf{Variables} & \textbf{Conditions} \\
\hline
Decomposition method 1 & $\kappa^{(1)}_{1}=0.3$; $\kappa^{(2)}_{1}=0.6$ \\
Decomposition method 2 & $\kappa^{(1)}_{2}=0.3$; $\kappa^{(2)}_{2}=0.6$ \\
\hline
\hline
\multicolumn{2}{c}{\textbf{Other parameters}}\\
\hline
\hline
\textbf{Variables} & \textbf{Conditions} \\
\hline
Residual covariance & $\theta^{(k)[xy]}_{\epsilon}=0.3\times\sqrt{\theta^{(k)[x]}_{\epsilon}\times\theta^{(k)[y]}_{\epsilon}}$ \\
\hline
\hline
\end{tabular}
\label{tbl:simu_design}
\begin{tablenotes}
\small
\item[a] {Relative rate-of-change is defined as the absolute rate-of-change over the shape factor (i.e., the slope in the first time interval).} \\
\item[b] {The correlation between the second-type time-invariant covariate and the trait feature is wet as $0.3$ so that they together explain $13\%$ and $26\%$ variance of growth factors of the longitudinal outcome in the first and second latent class, respectively.}
\end{tablenotes}
\end{threeparttable}}
\end{table}

\begin{table}
\centering
\resizebox{1.1\textwidth}{!}{
\begin{threeparttable}
\small
\setlength{\tabcolsep}{4pt}
\renewcommand{\arraystretch}{0.75}
\caption{Summary of Model Fit Information For the Growth Mixture Models\tnote{a}}
\begin{tabular}{lrrrrrrrr}
\hline
\hline
\textbf{Model} & \textbf{-2ll} & \textbf{AIC}  & \textbf{BIC}  & \textbf{\# of Para.} & \textbf{Class 1 Res.} & \textbf{Class 2 Res.} & \textbf{Class 3 Res.} & \textbf{Class 4 Res.} \\
\hline
\multicolumn{9}{c}{\textbf{No covariates}}\\
\hline
$1$ class & $31444.49$ & $31466.49$ & $31512.85$ & $11$ & $33.58$ & -- & -- & -- \\
$2$ classes & $31430.06$ & $31476.06$ & $31573.00$ & $23$ & $24.10$ & $33.89$ & -- & -- \\
$3$ classes & $31139.79$ & $31209.79$ & $31357.30$ & $35$ & $23.91$ & $32.69$ & $33.00$ & -- \\
$4$ classes & $31099.40$ & $31193.40$ & $31391.49$ & $47$ & $24.18$ & $32.03$ & $31.31$ & $33.27$ \\
\hline
\hline
\multicolumn{9}{c}{\textbf{Two first-type TICs, one second-type TIC, and baseline TVC\tnote{b}}}\\
\hline
$3$ classes & $33417.48$ & $33561.48$ & $33864.93$ & $72$ & $25.99$ & $32.87$ & $33.82$ & -- \\
\hline
\hline
\multicolumn{9}{c}{\textbf{Two first-type TICs, one second-type TIC, and a decomposed TVC (interval-specific slopes)}}\\
\hline
$1$ class & $41522.43$ & $41592.43$ & $41739.94$ & $35$ & $31.23$ & -- & -- & -- \\
$2$ classes & $40701.18$ & $40847.18$ & $41235.04$ & $73$ & $29.49$ & $32.59$ & -- & -- \\
$3$ classes & $40286.35$ & $40508.35$ & $41130.05$ & $111$ & $30.06$ & $28.64$ & $34.57$ & -- \\
\hline
\hline
\multicolumn{9}{c}{\textbf{Two first-type TICs, one second-type TIC, and a decomposed TVC (interval-specific changes)}}\\
\hline
$1$ class & $41502.81$ & $41572.81$ & $41720.32$ & $35$ & $31.06$ & -- & -- & -- \\
$2$ classes & $40691.97$ & $40837.97$ & $41225.83$ & $73$ & $29.20$ & $32.53$ & -- & -- \\
$3$ classes & $40266.80$ & $40488.80$ & $41110.50$ & $111$ & $29.55$ & $28.61$ & $34.27$ & -- \\
\hline
\hline
\end{tabular}
\label{tbl:info}
\begin{tablenotes}
\small
\item[a] In this application, the growth mixture models take the bilinear spline growth model with an unknown fixed knot.\\
\item[b] In this case, the growth mixture model is built with two second-type TICs: the standardized baseline teacher-reported inhibitory control and the standardized baseline reading scores.
\end{tablenotes}
\end{threeparttable}}
\end{table}

\begin{table}
\centering
\resizebox{1.00\textwidth}{!}{
\begin{threeparttable}
\small
\setlength{\tabcolsep}{4pt}
\renewcommand{\arraystretch}{0.75}
\caption{Estimates of Growth Mixture Model\tnote{a} with Two First-type Covariates, One Second-type Covariate, and a Decomposed TVC with Interval-specific Slopes}
\begin{tabular}{lrrrrrr}
\hline
\hline
& \multicolumn{2}{c}{Class 1} & \multicolumn{2}{c}{Class 2} & \multicolumn{2}{c}{Class 3} \\
\hline
Para. & Estimate (SE) & p-value & Estimate (SE) & p-value & Estimate (SE) & p-value \\
\hline
\multicolumn{7}{c}{Parameters related to mathematics development (longitudinal outcome)} \\
\hline
$\mu^{[y]}_{\eta_{0}}$ & $18.7756$ ($0.9895$) & $<0.0001^{\ast}$ & $24.553$ ($0.7401$) & $<0.0001^{\ast}$ & $33.077$ ($1.1048$) & $<0.0001^{\ast}$ \\
$\mu^{[y]}_{\eta_{1}}$ & $17.558$ ($0.4026$) & $<0.0001^{\ast}$ & $20.3812$ ($0.3353$) & $<0.0001^{\ast}$ & $22.5925$ ($0.3132$) & $<0.0001^{\ast}$ \\
$\mu^{[y]}_{\eta_{2}}$ & $8.3875$ ($0.7012$) & $<0.0001^{\ast}$ & $8.7642$ ($0.3426$) & $<0.0001^{\ast}$ & $8.3685$ ($0.3366$) & $<0.0001^{\ast}$ \\
$\gamma$ & $9.1221$ ($0.1036$) & $<0.0001^{\ast}$ & $8.5924$ ($0.0604$) & $<0.0001^{\ast}$ & $8.5063$ ($0.0519$) & $<0.0001^{\ast}$ \\
$\psi^{[y]}_{00}$ & $53.0067$ ($10.6947$) & $<0.0001^{\ast}$ & $48.7222$ ($7.9945$) & $<0.0001^{\ast}$ & $96.7551$ ($15.4978$) & $<0.0001^{\ast}$ \\
$\psi^{[y]}_{01}$ & $0.4931$ ($3.2883$) & $0.8808$ & $-0.2714$ ($2.5574$) & $0.9155$ & $-8.6971$ ($3.6204$) & $0.0163^{\ast}$ \\
$\psi^{[y]}_{02}$ & $-9.9621$ ($4.9921$) & $0.0460^{\ast}$ & $-7.6693$ ($2.3604$) & $0.0012^{\ast}$ & $-12.8962$ ($3.5548$) & $0.0003^{\ast}$ \\
$\psi^{[y]}_{11}$ & $9.6704$ ($1.8173$) & $<0.0001^{\ast}$ & $7.4731$ ($1.5655$) & $<0.0001^{\ast}$ & $6.2249$ ($1.4986$) & $<0.0001^{\ast}$ \\
$\psi^{[y]}_{12}$ & $-3.4295$ ($2.2594$) & $0.1290$ & $-1.0474$ ($1.0620$) & $0.3240$ & $-1.3718$ ($0.9865$) & $0.1643$ \\
$\psi^{[y]}_{12}$ & $20.1641$ ($4.8916$) & $<0.0001^{\ast}$ & $3.8244$ ($1.4182$) & $0.0070^{\ast}$ & $3.1989$ ($1.3570$) & $0.0184^{\ast}$ \\
$\theta^{[y]}$ & $30.0572$ ($1.6729$) & $<0.0001^{\ast}$ & $28.6395$ ($1.3297$) & $<0.0001^{\ast}$ & $34.5703$ ($1.5517$) & $<0.0001^{\ast}$ \\
\hline
\multicolumn{7}{c}{Parameters related to standardized reading ability (time-varying covariate)} \\
\hline
$\mu^{[x]}_{\eta_{0}}$ & $-0.6677$ ($0.0545$) & $<0.0001^{\ast}$ & $-0.2596$ ($0.0440$) & $<0.0001^{\ast}$ & $0.8104$ ($0.0999$) & $<0.0001^{\ast}$ \\
$\mu^{[x]}_{\eta_{1}}$ & $1.5141$ ($0.1007$) & $<0.0001^{\ast}$ & $1.9630$ ($0.0836$) & $<0.0001^{\ast}$ & $2.5522$ ($0.1135$) & $<0.0001^{\ast}$ \\
$\psi^{[x]}_{00}$ & $0.1495$ ($0.0329$) & $<0.0001^{\ast}$ & $0.1692$ ($0.0291$) & $<0.0001^{\ast}$ & $1.3213$ ($0.1560$) & $<0.0001^{\ast}$ \\
$\psi^{[x]}_{01}$ & $0.0261$ ($0.0114$) & $0.0218^{\ast}$ & $-0.0760$ ($0.0145$) & $<0.0001^{\ast}$ & $-0.4897$ ($0.0669$) & $<0.0001^{\ast}$ \\
$\psi^{[x]}_{11}$ & $0.0474$ ($0.0099$) & $<0.0001^{\ast}$ & $0.0747$ ($0.0115$) & $<0.0001^{\ast}$ & $0.2200$ ($0.0343$) & $<0.0001^{\ast}$ \\
$\gamma_{2}$ & $0.4618$ ($0.0803$) & $<0.0001^{\ast}$ & $0.6471$ ($0.0597$) & $<0.0001^{\ast}$ & $0.7956$ ($0.0611$) & $<0.0001^{\ast}$ \\
$\gamma_{3}$ & $1.3097$ ($0.1034$) & $<0.0001^{\ast}$ & $1.6584$ ($0.0855$) & $<0.0001^{\ast}$ & $1.0160$ ($0.0655$) & $<0.0001^{\ast}$ \\
$\gamma_{4}$ & $0.7101$ ($0.0808$) & $<0.0001^{\ast}$ & $0.5197$ ($0.0486$) & $<0.0001^{\ast}$ & $0.3890$ ($0.0449$) & $<0.0001^{\ast}$ \\
$\gamma_{5}$ & $1.2755$ ($0.0991$) & $<0.0001^{\ast}$ & $0.7027$ ($0.0511$) & $<0.0001^{\ast}$ & $0.5052$ ($0.0443$) & $<0.0001^{\ast}$ \\
$\gamma_{6}$ & $0.6534$ ($0.0606$) & $<0.0001^{\ast}$ & $0.3197$ ($0.0268$) & $<0.0001^{\ast}$ & $0.1484$ ($0.0237$) & $<0.0001^{\ast}$ \\
$\gamma_{7}$ & $0.4748$ ($0.0491$) & $<0.0001^{\ast}$ & $0.3144$ ($0.0264$) & $<0.0001^{\ast}$ & $0.2552$ ($0.0242$) & $<0.0001^{\ast}$ \\
$\gamma_{8}$ & $0.3621$ ($0.0440$) & $<0.0001^{\ast}$ & $0.2706$ ($0.0247$) & $<0.0001^{\ast}$ & $0.1949$ ($0.0236$) & $<0.0001^{\ast}$ \\
$\theta^{[x]}$ & $0.1951$ ($0.0102$) & $<0.0001^{\ast}$ & $0.1633$ ($0.0073$) & $<0.0001^{\ast}$ & $0.2751$ ($0.0115$) & $<0.0001^{\ast}$ \\
\hline
\multicolumn{7}{c}{Parameters related to baseline inhibitory control (TIC)} \\
\hline
$\mu_{x}$ & $-0.4372$ ($0.0962$) & $<0.0001^{\ast}$ & $0.0983$ ($0.0695$) & $0.1574$ & $0.1920$ ($0.0782$) & $0.0141^{\ast}$ \\
$\phi_{x}$ & $1.0248$ ($0.1331$) & $<0.0001^{\ast}$ & $0.7906$ ($0.0854$) & $<0.0001^{\ast}$ & $1.0243$ ($0.1103$) & $<0.0001^{\ast}$ \\
\hline
\multicolumn{7}{c}{Correlation between TIC and baseline TVC} \\
\hline
$\rho_{\text{BL}}$ & $0.1297$ ($0.0470$) & $0.0058^{\ast}$ & $0.0636$ ($0.0256$) & $0.0132^{\ast}$ & $0.0804$ ($0.0442$) & $0.0691$ \\
\hline
\multicolumn{7}{c}{Effect of the TIC} \\
\hline
$\beta_{TIC_{0}}$ & $1.0024$ ($0.9225$) & $0.2772$ & $0.4919$ ($0.7445$) & $0.5088$ & $0.6531$ ($0.8849$) & $0.4605$ \\
$\beta_{TIC_{1}}$ & $0.2261$ ($0.4058$) & $0.5774$ & $1.2357$ ($0.3086$) & $0.0001^{\ast}$ & $-0.2836$ ($0.2746$) & $0.3017$ \\
$\beta_{TIC_{2}}$ & $0.2934$ ($0.6283$) & $0.6406$ & $-1.0224$ ($0.3122$) & $0.0011^{\ast}$ & $0.3121$ ($0.2668$) & $0.2421$ \\
\hline
\multicolumn{7}{c}{Trait effect of the TVC} \\
\hline
$\beta_{TVC_{0}}$ & $11.1670$ ($2.9495$) & $0.0002^{\ast}$ & $6.4729$ ($2.0437$) & $0.0015^{\ast}$ & $6.2934$ ($0.8232$) & $<0.0001^{\ast}$ \\
$\beta_{TVC_{1}}$ & $1.7818$ ($1.2583$) & $0.1568$ & $-3.9523$ ($0.9389$) & $<0.0001^{\ast}$ & $-0.9547$ ($0.2504$) & $0.0001^{\ast}$ \\
$\beta_{TVC_{2}}$ & $1.4842$ ($2.0238$) & $0.4633$ & $0.3563$ ($0.8618$) & $0.6793$ & $-0.4979$ ($0.2340$) & $0.0333^{\ast}$ \\
\hline
\multicolumn{7}{c}{State effect of the TVC} \\
\hline
$\kappa_{1}$ & $3.3770$ ($0.3800$) & $<0.0001^{\ast}$ & $1.9573$ ($0.1838$) & $<0.0001^{\ast}$ & $1.3531$ ($0.2037$) & $<0.0001^{\ast}$ \\
\hline
\multicolumn{7}{c}{Covariance between residual of longitudinal outcome and that of TVC} \\
\hline
$\theta^{[xy]}$ & $0.3105$ ($0.1011$) & $0.0021^{\ast}$ & $0.3045$ ($0.0682$) & $<0.0001^{\ast}$ & $0.2052$ ($0.0926$) & $0.0268^{\ast}$ \\
\hline
\multicolumn{7}{c}{Logistic coefficients} \\
\hline
$\beta_{0}$ & --\tnote{b} & --\tnote{b} & $0.5748$ ($0.1656$) & $0.0005^{\ast}$ & $0.3931$ ($0.1594$) & $0.0137^{\ast}$ \\
$\beta_{1}$ & --\tnote{b} & --\tnote{b} & $0.6208$ ($0.1724$) & $0.0003^{\ast}$ & $0.5523$ ($0.1703$) & $0.0012^{\ast}$ \\
$\beta_{2}$ & --\tnote{b} & --\tnote{b} & $-0.1343$ ($0.1703$) & $0.4303$ & $0.6617$ ($0.1774$) & $0.0002^{\ast}$ \\
\hline
\hline
\end{tabular}
\label{tbl:TVCslp_est}
\begin{tablenotes}
\small
\item[a] In this application, the growth mixture models take the bilinear spline growth model with an unknown fixed knot.\\
\item[b] -- indicates the estimates and p-values are unavailable as the first latent class was set as the reference group in the Application section.\\
\item {$^{\ast}$ indicates statistical significance at $0.05$ level.}
\end{tablenotes}
\end{threeparttable}}
\end{table}

\begin{table}
\centering
\resizebox{1.00\textwidth}{!}{
\begin{threeparttable}
\small
\setlength{\tabcolsep}{4pt}
\renewcommand{\arraystretch}{0.75}
\caption{Estimates of Growth Mixture Model\tnote{a} with Two First-type Covariates, One Second-type Covariate, and a Decomposed TVC with Interval-specific Changes}
\begin{tabular}{lrrrrrr}
\hline
\hline
& \multicolumn{2}{c}{Class 1} & \multicolumn{2}{c}{Class 2} & \multicolumn{2}{c}{Class 3} \\
\hline
Para. & Estimate (SE) & p-value & Estimate (SE) & p-value & Estimate (SE) & p-value \\
\hline
\multicolumn{7}{c}{Parameters related to mathematics development (longitudinal outcome)} \\
\hline
$\mu^{[y]}_{\eta_{0}}$ & $19.3405$ ($0.9749$) & $<0.0001^{\ast}$ & $24.7378$ ($0.7257$) & $<0.0001^{\ast}$ & $33.1276$ ($1.0985$) & $<0.0001^{\ast}$ \\
$\mu^{[y]}_{\eta_{1}}$ & $16.7201$ ($0.5079$) & $<0.0001^{\ast}$ & $20.4099$ ($0.3466$) & $<0.0001^{\ast}$ & $22.5323$ ($0.3174$) & $<0.0001^{\ast}$ \\
$\mu^{[y]}_{\eta_{2}}$ & $10.1463$ ($0.6082$) & $<0.0001^{\ast}$ & $9.0398$ ($0.3300$) & $<0.0001^{\ast}$ & $8.2919$ ($0.3354$) & $<0.0001^{\ast}$ \\
$\gamma$ & $8.7558$ ($0.0087$) & $<0.0001^{\ast}$ & $8.4340$ ($0.0591$) & $<0.0001^{\ast}$ & $8.4759$ ($0.0502$) & $<0.0001^{\ast}$ \\
$\psi^{[y]}_{00}$ & $50.3895$ ($10.8032$) & $<0.0001^{\ast}$ & $45.9351$ ($7.8522$) & $<0.0001^{\ast}$ & $97.4480$ ($14.795$) & $<0.0001^{\ast}$ \\
$\psi^{[y]}_{01}$ & $-0.9228$ ($3.5817$) & $0.7967$ & $0.1694$ ($2.6114$) & $0.9483$ & $-9.1609$ ($3.6685$) & $0.0125^{\ast}$ \\
$\psi^{[y]}_{02}$ & $-4.1603$ ($4.2215$) & $0.3244$ & $-6.9253$ ($2.1740$) & $0.0014^{\ast}$ & $-12.5128$ ($3.1822$) & $0.0001^{\ast}$ \\
$\psi^{[y]}_{11}$ & $10.6442$ ($2.1160$) & $<0.0001^{\ast}$ & $7.9729$ ($1.5977$) & $<0.0001^{\ast}$ & $6.4247$ ($1.4745$) & $<0.0001^{\ast}$ \\
$\psi^{[y]}_{12}$ & $-2.4164$ ($1.9739$) & $0.2209$ & $-0.9180$ ($1.0283$) & $0.3720$ & $-1.3818$ ($0.9792$) & $0.1582$ \\
$\psi^{[y]}_{12}$ & $15.4057$ ($3.6889$) & $<0.0001^{\ast}$ & $3.5322$ ($1.2021$) & $0.0033^{\ast}$ & $3.1684$ ($1.2320$) & $0.0101^{\ast}$ \\
$\theta^{[y]}$ & $29.5455$ ($1.6467$) & $<0.0001^{\ast}$ & $28.6058$ ($1.3182$) & $<0.0001^{\ast}$ & $34.2691$ ($1.5335$) & $<0.0001^{\ast}$ \\
\hline
\multicolumn{7}{c}{Parameters related to standardized reading ability (time-varying covariate)} \\
\hline
$\mu^{[x]}_{\eta_{0}}$ & $-0.6487$ ($0.0552$) & $<0.0001^{\ast}$ & $-0.2629$ ($0.0439$) & $<0.0001^{\ast}$ & $0.8114$ ($0.0997$) & $<0.0001^{\ast}$ \\
$\mu^{[x]}_{\eta_{1}}$ & $1.4232$ ($0.1026$) & $<0.0001^{\ast}$ & $1.9700$ ($0.0840$) & $<0.0001^{\ast}$ & $2.5409$ ($0.1133$) & $<0.0001^{\ast}$ \\
$\psi^{[x]}_{00}$ & $0.1502$ ($0.0329$) & $<0.0001^{\ast}$ & $0.1712$ ($0.0291$) & $<0.0001^{\ast}$ & $1.3227$ ($0.1559$) & $<0.0001^{\ast}$ \\
$\psi^{[x]}_{01}$ & $0.0255$ ($0.0107$) & $0.0167^{\ast}$ & $-0.0751$ ($0.0145$) & $<0.0001^{\ast}$ & $-0.4885$ ($0.0668$) & $<0.0001^{\ast}$ \\
$\psi^{[x]}_{11}$ & $0.0426$ ($0.0091$) & $<0.0001^{\ast}$ & $0.0745$ ($0.0115$) & $<0.0001^{\ast}$ & $0.2185$ ($0.0341$) & $<0.0001^{\ast}$ \\
$\gamma_{2}$ & $0.5455$ ($0.0891$) & $<0.0001^{\ast}$ & $0.6474$ ($0.0598$) & $<0.0001^{\ast}$ & $0.8057$ ($0.0618$) & $<0.0001^{\ast}$ \\
$\gamma_{3}$ & $1.3237$ ($0.1035$) & $<0.0001^{\ast}$ & $1.6397$ ($0.0845$) & $<0.0001^{\ast}$ & $1.0163$ ($0.0650$) & $<0.0001^{\ast}$ \\
$\gamma_{4}$ & $0.8460$ ($0.0981$) & $<0.0001^{\ast}$ & $0.5236$ ($0.0489$) & $<0.0001^{\ast}$ & $0.3982$ ($0.0451$) & $<0.0001^{\ast}$ \\
$\gamma_{5}$ & $1.3214$ ($0.1022$) & $<0.0001^{\ast}$ & $0.6910$ ($0.0504$) & $<0.0001^{\ast}$ & $0.4981$ ($0.0437$) & $<0.0001^{\ast}$ \\
$\gamma_{6}$ & $0.6833$ ($0.0630$) & $<0.0001^{\ast}$ & $0.3233$ ($0.0262$) & $<0.0001^{\ast}$ & $0.1535$ ($0.0232$) & $<0.0001^{\ast}$ \\
$\gamma_{7}$ & $0.5505$ ($0.0516$) & $<0.0001^{\ast}$ & $0.3214$ ($0.0257$) & $<0.0001^{\ast}$ & $0.2575$ ($0.0235$) & $<0.0001^{\ast}$ \\
$\gamma_{8}$ & $0.3431$ ($0.0411$) & $<0.0001^{\ast}$ & $0.2584$ ($0.0242$) & $<0.0001^{\ast}$ & $0.1899$ ($0.0232$) & $<0.0001^{\ast}$ \\
$\theta^{[x]}$ & $0.1953$ ($0.0103$) & $<0.0001^{\ast}$ & $0.1636$ ($0.0072$) & $<0.0001^{\ast}$ & $0.2750$ ($0.0115$) & $<0.0001^{\ast}$ \\
\hline
\multicolumn{7}{c}{Parameters related to baseline inhibitory control (TIC)} \\
\hline
$\mu_{x}$ & $-0.4493$ ($0.0972$) & $<0.0001^{\ast}$ & $0.1004$ ($0.0691$) & $0.1464$ & $0.1948$ ($0.0778$) & $0.0123^{\ast}$ \\
$\phi_{x}$ & $1.0299$ ($0.1344$) & $<0.0001^{\ast}$ & $0.7852$ ($0.0846$) & $<0.0001^{\ast}$ & $1.0206$ ($0.1096$) & $<0.0001^{\ast}$ \\
\hline
\multicolumn{7}{c}{Correlation between TIC and baseline TVC} \\
\hline
$\rho_{\text{BL}}$ & $0.1270$ ($0.0473$) & $0.0072$ & $0.0621$ ($0.0258$) & $0.0159^{\ast}$ & $0.0786$ ($0.0441$) & $0.0744$ \\
\hline
\multicolumn{7}{c}{Effect of the TIC} \\
\hline
$\beta_{TIC_{0}}$ & $0.8214$ ($0.8943$) & $0.3584$ & $0.3570$ ($0.7310$) & $0.6253$ & $0.6987$ ($0.8821$) & $0.4283$ \\
$\beta_{TIC_{1}}$ & $0.2673$ ($0.3920$) & $0.4953$ & $1.2728$ ($0.3242$) & $0.0001^{\ast}$ & $-0.3036$ ($0.2758$) & $0.2708$ \\
$\beta_{TIC_{2}}$ & $0.3293$ ($0.5021$) & $0.5119$ & $-0.8917$ ($0.2941$) & $0.0024^{\ast}$ & $0.3028$ ($0.2600$) & $0.2442$ \\
\hline
\multicolumn{7}{c}{Trait effect of the TVC} \\
\hline
$\beta_{TVC_{0}}$ & $11.5206$ ($2.8698$) & $0.0001^{\ast}$ & $6.7852$ ($1.9983$) & $0.0007^{\ast}$ & $6.3065$ ($0.8305$) & $<0.0001^{\ast}$ \\
$\beta_{TVC_{1}}$ & $1.8332$ ($1.3080$) & $0.1611$ & $-4.0325$ ($0.9349$) & $<0.0001^{\ast}$ & $-0.9312$ ($0.2537$) & $0.0002^{\ast}$ \\
$\beta_{TVC_{2}}$ & $0.9973$ ($1.8452$) & $0.5889$ & $0.3633$ ($0.7891$) & $0.6453$ & $-0.4893$ ($0.2327$) & $0.0355^{\ast}$ \\
\hline
\multicolumn{7}{c}{State effect of the TVC} \\
\hline
$\kappa_{2}$ & $8.1299$ ($1.0387$) & $<0.0001^{\ast}$ & $3.7097$ ($0.3593$) & $<0.0001^{\ast}$ & $2.7577$ ($0.3945$) & $<0.0001^{\ast}$ \\
\hline
\multicolumn{7}{c}{Covariance between residual of longitudinal outcome and that of TVC} \\
\hline
$\theta^{[xy]}$ & $0.2667$ ($0.1007$) & $0.0081^{\ast}$ & $0.3012$ ($0.0680$) & $<0.0001^{\ast}$ & $0.2057$ ($0.0923$) & $0.0259^{\ast}$ \\
\hline
\multicolumn{7}{c}{Logistic coefficients} \\
\hline
$\beta_{0}$ & --\tnote{b} & --\tnote{b} & $0.5897$ ($0.1618$) & $0.0003^{\ast}$ & $0.4004$ ($0.1581$) & $0.0113^{\ast}$ \\
$\beta_{1}$ & --\tnote{b} & --\tnote{b} & $0.6273$ ($0.1717$) & $0.0003^{\ast}$ & $0.5641$ ($0.1710$) & $0.0010^{\ast}$ \\
$\beta_{2}$ & --\tnote{b} & --\tnote{b} & $-0.1475$ ($0.1691$) & $0.3831$ & $0.6526$ ($0.1773$) & $0.0002^{\ast}$ \\
\hline
\hline
\end{tabular}
\label{tbl:TVCchg_est}
\begin{tablenotes}
\small
\item[a] In this application, the growth mixture models take the bilinear spline growth model with an unknown fixed knot.\\
\item[b] -- indicates the estimates and p-values are unavailable as the first latent class was set as the reference group in the Application section.\\
\item {$^{\ast}$ indicates statistical significance at $0.05$ level.}
\end{tablenotes}
\end{threeparttable}}
\end{table}

\end{document}